\documentclass[aps,preprint]{revtex4}

\usepackage[dvips]{graphicx}

\newcommand{\ch}{{\cal H}}
\newcommand{\ce}{{\cal E}}

\begin{document}
\date{\today}
\title{
Open quantum systems and Dicke superradiance
}

\author{Hichem Eleuch$^{1,2}$\footnote{email: hichemeleuch@yahoo.fr} and 
Ingrid Rotter$^{3}$\footnote{email: rotter@pks.mpg.de}}

\address{
$^1$Universite de Montreal, C.P. 6128, 
Montreal (QC), H3C 3J7 Canada}
\address{
$^2$Department of Physics, McGill University, Montreal, Canada  H3A  2T8}
\address{
$^3$Max Planck Institute for the Physics of Complex Systems,
D-01187 Dresden, Germany }

\begin{abstract}

We study generic features of  open quantum systems embedded 
into a continuum of scattering wavefunctions and compare them with
results discussed in optics. 
A dynamical phase transition  may appear at high level density in a
many-level system and also in a two-level system if the  coupling
$W$ to the environment is complex and sufficiently large.
Here nonlinearities occur.
When $W_{ij}$ is imaginary, two singular (exceptional) points 
may exist. In the parameter range between these two points, width bifurcation 
occurs as function of a certain external parameter. 
A unitary representation of the S-matrix allows to calculate 
the cross section for a two-level system,
including at the exceptional point (double pole of the S-matrix). The
results obtained for the transition of level repulsion at small (real)
$W_{ij}$ to width bifurcation at large (imaginary) $W_{ij}$ show qualitatively
the same features that are observed experimentally in the transition
from Autler-Townes splitting to electromagnetically induced
transparency in optics. Fermi's golden rule holds only below the
dynamical phase transition while it passes into an anti-golden rule beyond
this transition. The results are generic and can be applied to
the response of a complex open 
quantum system to the action of an external field (environment). 
They may be considered as a guideline for
engineering and manipulating quantum systems in such a way that they
can be used for applications with special requirements.

\end{abstract}

\pacs{\bf }

\maketitle

\section{Introduction}
\label{intr}

Recently, dynamical phase transitions (DPTs) are considered in 
different open quantum systems. They appear mostly at high level density and 
are observed experimentally as well as
theoretically by using different approaches. Common to all of them is
a very robust
spectroscopic redistribution that takes place in the system in a
critical region of a certain control parameter. As a result of the
spectroscopic redistribution,  short-lived states appear together
with long-lived ones  (called usually line width bifurcation) which all have lost
their spectroscopic relation to the original states in the 
subcritical parameter region. Mathematically, the
spectroscopic redistribution can be traced back to the existence of singular
points in the continuum (called usually  exceptional points
(EPs)).   For details see  Appendix \ref{ap1}.

The DPTs observed in many-body open quantum systems 
show features which are similar to the Dicke superradiance known in optics
\cite{dicke}.  This has been shown some years ago \cite{soze}\,:
the DPT observed in a many-body open quantum system 
\cite{klro} is analogue to the formation of the superradiant 
Dicke state in optics. The simple model used in this study is based on 
statistical assumptions and can, of course, not
explain the sensitive dependence of the system properties on the
variation of external parameters. It shows however the meaning of the 
imaginary part of the coupling term between system and environment 
for the formation of the so-called superradiant (short-lived) and
subradiant (long-lived) states in a many-body open 
quantum system. The imaginary part of the coupling term causes the
Hamiltonian to be non-Hermitian. 

The similarity between many-body open quantum systems  and optics is
observed also in other papers.
Cooperative spontaneous emission is considered as a many-body
eigenvalue problem in \cite{svid}. As a result, some states decay much
faster than the single-atom decay rate, while other states are trapped
and undergo very slow decay. When the size of the atomic cloud is
small compared with the radiation wavelength, 
the eigenvalues have a large imaginary
part which corresponds to a large frequency shift of the emitted radiation.
The collective Lamb shift in single photon Dicke superradiance 
is formulated theoretically  in \cite{scully3} and observed
experimentally \cite{roehlsberger}, see also \cite{scully}. 
Furthermore, Fermi's golden rule does not adequately describe
superradiance in optics  \cite{scully1}, what is along the lines of 
observations in open quantum systems 
beyond the DPT \cite{past1,past2}. This fact suggests to consider 
the possibility that a DPT of the type discussed above for a many-body
quantum system appears also in optics.

Under critical conditions, line width bifurcation is found in
different many-body open quantum systems both theoretically and
experimentally. Some of them are mentioned in the review \cite{top}. 
Examples related to optics are the studies on the Dicke-model quantum
phase transition in an optical cavity \cite{diss2} and further studies
of the open-system Dicke model \cite{domo1}. Interesting is the
observation that a gas of laser-driven cold atoms, scattering light
into a cavity, produces a phase transition between the homogeneous
spatial distribution and a regular periodic pattern \cite{domo3}. 
This phase transition is related to selforganization \cite{domo2}.

Recently, the superradiant states in optical devices are
studied experimentally in many papers. 
They appear coherently when the single parts (states)
of the system come into contact with one another, e.g. \cite{kaiser1,kaiserfdp}.
Further,  long-lived subradiant states \cite{kaiser2} and superradiant 
forward scattering \cite{kaiser3} are observed. Correlated spontaneous
emission from an ensemble of $N$ identical two-level atoms is
considered in  \cite{scully2}.
The transition between Autler-Townes splitting \cite{autler} and
electromagnetically induced transparency  shows features \cite{giner}
which are similar to level repulsion and width bifurcation observed in a 
many-body system (Fig. 9 in \cite{mudiisro}).
A similar effect is observed in the electromagnetically induced
transparency on a single artificial atom \cite{abdu}.
Electromagnetically induced transparency is studied 
experimentally including its
dependence on the sample geometry \cite{roehlsbergerfdp}.
Also superradiance of quantum dots is experimentally found 
\cite{scheibner}. An enhancement of photon
intensity in forced coupled quantum wells inside a semiconductor microcavity
is discussed in \cite{prasad}.
In a recent paper \cite{zhang}, the non-Hermitian formalism is used to
design a plasmonic system for spatially manipulating light on the
nanoscale, and the selective excitation of each individual element in
the array is experimentally demonstrated. Nearly complete
suppression of dynamical tunneling in asymmetric resonant cavities  
is  possible due to the phenomenon of avoided resonance crossing
as shown recently \cite{tunnel}.

The non-Hermitian Hamiltonian in quantum physics is an expression of the
fact that the system considered is really open\,: it is embedded into
the continuum of scattering wavefunctions which always exists,
and its properties are influenced by the coupling to this environment.
The interaction of the states via the environment
is  unimportant at low level density where it can be
described by perturbation theory (if necessary). It becomes, however, decisive
in the regime of overlapping resonances \cite{top}. Here, the eigenvalues
of the non-Hermitian Hamiltonian
${\cal H}$ avoid crossing in energy ({\it level repulsion})
and bifurcate in time ({\it width bifurcation}). 
This effect is caused mathematically by the
existence of singular points (mostly called EPs) which introduce
nonlinear effects into the basic equations. At an EP,
two eigenvalues of  ${\cal H}$ coalesce, the two corresponding 
eigenfunctions are not orthogonal to one another but linearly
dependent, and the phases of the eigenfunctions are not rigid.
The existence of EPs 
has been proven experimentally by means of microwave billiards \cite{demb1}.

The aim of the present paper is to simulate a DPT 
in an open quantum system by using a schematical (toy) model.  
The eigenvalues $\ce_i$ of a non-Hermitian Hamilton operator $\ch$
are traced  as  function of the distance to neighboring  eigenstates. 
The non-diagonal matrix elements $\omega_{ij}$ 
of $\ch$ simulate the coupling between the states $i$ and $j$
via the environment (continuum of scattering wavefunctions or another
external field) and are complex, see Appendix \ref{ap2}. 
Hence, the eigenvalues  $\ce_{i,j}$ of $\ch$ contain the
influence of the environment onto the states of the system\,:
level repulsion is caused by 
Re($\omega_{ij}$)  and width bifurcation by Im($\omega_{ij}$). 
In realistic systems,  Re$(\omega_{ij}) \gg {\rm Im}(\omega_{ij}) $ 
far from EPs while 
Im$(\omega_{ij}) \gg {\rm Re}(\omega_{ij}) $  near to an EP. 
Here, the eigenvalue trajectories avoid crossing and a
transition from level repulsion in energy to width bifurcation takes place.

The paper is organized as follows.  Sect. \ref{eigen}
sketches the  formalism used in the present
paper. The basis is  the $2\times 2$ Hamiltonian matrix $\ch^{(2)}$
which describes two states of the system and their
coupling to and via the environment (continuum of scattering
wavefunctions or an external field). The physical meaning of all
matrix elements is explained in the Appendix \ref{ap2} by using  
the non-Hermitian Hamilton operator (\ref{form5})
as well as the equations (\ref{form11})   and (\ref{form12}) 
which are characteristic of an open quantum system. Furthermore, 
the basic expressions such as  EP,
biorthogonality, phase rigidity, level repulsion, width bifurcation,
mixing of the eigenfunctions  are defined and an expression for the
$S$ matrix is given. In Sect. \ref{more}, the corresponding 
Hamiltonian matrix $\ch^{(N)}$ for an $N$
level system is written down and the physical meaning of the matrix elements
is discussed. 

In the following sections, some numerical results are given. 
The eigenvalues ${\cal E}_i$ and
eigenfunctions $\Phi_i$ of the $2\times 2$ Hamiltonian    $\ch^{(2)}$
with $N=2$ levels are given in Sect. \ref{num1} while those of  $\ch^{(N)}$
for $N=3$ and 4 are shown in 
Sect. \ref{num2}. In Sect. \ref{num3}, numerical results on the influence
of EPs onto the cross section  are given.

The results are discussed in Sect. \ref{disc}. Special attention is
devoted first to symmetries and nonlinearities around an EP and 
their role for an DPT. 
Secondly we point to a possible DPT in a two-level system
at strong coupling  to its environment and its relation to the transition
from Autler-Townes splitting to electromagnetically induced
transparency in optics. Furthermore,
the experimental observation of these different DPTs is discussed. 
The last section  \ref{concl} contains some conclusions drawn from the
results given in the main part of the paper.

\section{
Eigenvalues and eigenfunctions of a non-Hermitian operator
and nonlinearities around an exceptional point}
\label{eigen}
 
The general expression of the non-Hermitian Hamiltonian   $\ch$ 
describing an open many-body system is given in Appendix \ref{ap2}.  
For illustration, let us simulate it by considering
explicitly only two states of the system, i.e.  by the  
symmetric $2 \times 2$ Hamiltonian 
\begin{eqnarray}
\ch^{(2)} = \left(
\begin{array}{cc}
\epsilon_1 & \omega_{12} \\
\omega_{21} & \epsilon_2 
\end{array} \right) \, .
\label{int3}
 \end{eqnarray}
It is assumed here that the direct {\rm internal} interaction $V_{ij}$  
of the two states (see Appendix \ref{ap2} for its definition) 
is involved in the  energies  $\epsilon_i ~(i=1,2)$ of the two states. 
The $\omega_{12} = \omega_{21} $ describe the
{\rm external} interaction of the two states via the environment.
The eigenvalues of $\ch^{(2)}$ are 
\begin{eqnarray}
\ce_{1,2}&=&\frac{\epsilon_1 + \epsilon_2}{2} \pm Z \; ; \; \quad
Z=\frac{1}{2}\sqrt{(\epsilon_1 - \epsilon_2)^2 + 4 \omega_{12}^2} \; .
\label{int4}
\end{eqnarray}
As  function of a certain parameter, the  levels repel each other 
in energy according to the value
Re$(Z)$ while the  widths bifurcate corresponding to Im$(Z)$.
The two eigenvalue trajectories cross when $Z=0$, i.e. when
$(\epsilon_1 - \epsilon_2)/2\omega_{12} = \pm \, i $.
The crossing points are called mostly {\it exceptional points} (EPs) 
according to the definition given in \cite{kato}.
At these singular points, the two eigenvalues coalesce,
$\ce_1 ~=~ \ce_2 ~\equiv ~\ce_0 $. 
In the vicinity of the crossing points, the dependence of the 
eigenvalue trajectories on a certain parameter is more complicated
than far from them: the two levels approach each other in energy
and the widths become equal so that
Re$(\ce_1) \leftrightarrow {\rm Re}(\ce_2)$ and
Im$(\ce_1) \leftrightarrow {\rm Im}(\ce_2)$
at the crossing point. 

Generally, $\ch^{(2)}$ is a non-Hermitian operator, the unperturbed
energies $\epsilon_i $ and  the interaction $\omega_{ij}$
are complex, see  Appendix \ref{ap2}, Eqs. (\ref{form5}),  (\ref{form11}) 
and (\ref{form12}). The states  can decay, in general, and
the two eigenvalues  (\ref{int4}) can be written as \cite{comment2}
\begin{eqnarray} 
\ce_{1,2}= 
E_{1,2} + \frac{i}{2} ~\Gamma_{1,2}  \qquad ({\rm with} ~\Gamma_{1,2}
\le 0 ) \; .
\label{eiv1}
\end{eqnarray}
The widths $|\Gamma_i|$ are proportional to the inverse lifetimes
$\tau_i^{-1}$ of the  states, $i=1,2$.
The topological phase of the EP is half the Berry
phase (see Sect. 2.5 of \cite{top}). 
This theoretical result is proven experimentally
by means of a microwave cavity \cite{demb1}.

The eigenfunctions of the non-Hermitian Hamilton operator $\ch^{(2)}$,
Eq. (\ref{int3}), are biorthogonal,
\begin{eqnarray}
\langle \Phi_k^*|\Phi_l\rangle  = \delta_{k, l} \; .
\label{eif1}
\end{eqnarray}
From these equations follows \cite{top}
\begin{eqnarray}
\langle \Phi_k|\Phi_{k }\rangle & \equiv &  A_k \ge 1 
\label{eif2}\\
\langle \Phi_k|\Phi_{l\ne k}\rangle =- 
\langle \Phi_{l \ne k  }|\Phi_k\rangle & \equiv & B_k^l ~; 
~~~|B_k^l|\ge 0 \; .
\label{eif3}
\end{eqnarray}
The eigenfunctions $\Phi_i$ of  $\ch^{(2)}$ can be represented in the
set of basic wavefunctions $\Phi_i^0$ of the matrix (\ref{int3}) with
vanishing non-diagonal matrix elements $\omega_{ij}$,
\begin{eqnarray}
\Phi_i=\sum_{j=1}^{N=2} b_{ij} \Phi_j^0 \; . 
\label{mixb}
\end{eqnarray}
The $b_{ij}$ are normalized according to the biorthogonality relations
(\ref{eif1}) of the wavefunctions $\{\Phi_i\}$. They characterize the
mixing of the eigenfunctions of  $\ch^{(2)}$ due to the coupling 
of the states via the environment. 

At the crossing point  $A_k^{\rm (cr)} \to \infty,  
~|B_k^{l ~{\rm (cr)}}| \to \infty $ \cite{top}. 
The relation between the eigenfunctions 
$\Phi_1$ and $\Phi_2$ of (\ref{int3}) at the crossing point  is
\begin{eqnarray}
\Phi_1^{\rm cr} \to ~\pm ~i~\Phi_2^{\rm cr} \; ;
\quad \qquad \Phi_2^{\rm cr} \to
~\mp ~i~\Phi_1^{\rm cr}  
\label{eif5}
\end{eqnarray}  
according to analytical  as well as numerical studies, see Appendix of
\cite{fortsch1} and Sect. 2.5 of \cite{top}.
That means, the state $\Phi_1$ jumps at the EP via
the chiral state ~$\Phi_1\pm \, i\, \Phi_2$ ~to the state  ~$\pm\, i\, \Phi_2$
\cite{comment}.
 From (\ref{eif5}) follows\,:
\begin{verse}
(i) When  two levels are distant from one another,  their eigenfunctions
 are (almost) orthogonal,  
$\langle \Phi_k^* | \Phi_k \rangle   \approx
\langle \Phi_k | \Phi_k \rangle  \equiv A_k \approx 1 $.\\
(ii) When  two levels cross at the EP, 
their eigenfunctions are linearly
dependent according to (\ref{eif5}) and 
$\langle \Phi_k | \Phi_k \rangle \equiv A_k \to \infty $.\\
\end{verse}
These two relations show that the phases of the two eigenfunctions
relative to one another change when the crossing point is approached. 
This can be expressed quantitatively by defining the {\it phase
  rigidity} $r_k$ of the eigenfunctions $\Phi_k$,
\begin{eqnarray}
r_k ~\equiv ~\frac{\langle \Phi_k^* | \Phi_k \rangle}{\langle \Phi_k 
| \Phi_k \rangle} ~= ~A_k^{-1} \; . 
\label{eif11}
\end{eqnarray}
It holds $1 ~\ge ~r_k ~\ge ~0 $.  
The  non-rigidity $r_k$ of the phases of the eigenfunctions of $\ch^{(2)}$ 
follows also from the fact that $\langle\Phi_k^*|\Phi_k\rangle$
is a complex number (in difference to the norm
$\langle\Phi_k|\Phi_k\rangle$ which is a real number) 
such that the normalization condition
(\ref{eif1}) can be fulfilled only by the additional postulation 
Im$\langle\Phi_k^*|\Phi_k\rangle =0$ (what corresponds to a rotation). 

If $r_k<1$, an analytical expression for the eigenfunctions as 
function of a certain control parameter  can,
generally, not be obtained. An exception is the special case  
$\gamma_1 = \gamma_2$  for which
$Z=  \frac{1}{2} \sqrt{(e_1 - e_2)^2 + 4 \omega_{12}^2}$. 
In this case, the condition $Z=0$ can not be fulfilled if 
$\omega_{12} = x$ is real due to 
\begin{eqnarray}
(e_1 - e_2)^2 +4\, x^2 &>& 0  \, .
\label{int6a}
\end{eqnarray}
The EP can be found only by
analytical continuation into the continuum \cite{top,ro01} and  
the two states avoid crossing. This is analogue
to the avoided level crossings of discrete states. 

The condition $Z=0$ can be fulfilled however in the above case if
 $\omega_{12} = i\, x$ is imaginary,
\begin{eqnarray}
(e_1 - e_2)^2 -4\, x^2 &= &0 
~~\rightarrow ~~e_1 - e_2 =\pm \, 2\, x \; ,
\label{int6b}
\end{eqnarray}
and two EPs appear.  It holds further
\begin{eqnarray}
\label{int6c}
(e_1 - e_2)^2 >4\, x^2 &\rightarrow& ~Z ~\in ~\Re \\
\label{int6d}
(e_1 - e_2)^2 <4\, x^2 &\rightarrow&  ~Z ~\in ~\Im 
\end{eqnarray}
independent of any parameter dependence $e_i(a)$.
In the first case, the eigenvalues ${\cal E}_i = E_i+i/2\, \Gamma_i$ 
differ from the original values 
$\epsilon_i = e_i + i/2~\gamma_i$ by a contribution to
the energies ({\it level repulsion}) and in the second case by a contribution
to the widths ({\it width bifurcation}). 
The width bifurcation starts at one of the EPs and
becomes maximum in the middle between the two EPs.
This happens at the crossing point $e_1 = e_2$ where 
$\Delta \Gamma/2 \equiv |\Gamma_1/2 - \Gamma_2/2| = 4\, x$.  

If $r_k<1$, the Schr\"odinger equation contains nonlinear terms.
According to (\ref{int3}), the Schr\"odinger equation with the 
unperturbed operator $H_0\equiv H(\omega_{12} = 0)$ and a source term
arising from the interaction $\omega_{12}=\omega_{21}$ of the states
via the environment  reads 
\cite{ro01}
\begin{eqnarray}
\label{mix1}
(H_0  - \epsilon_n) ~| \Phi_n \rangle & = &
- \left(
\begin{array}{cc}
0 & \omega_{12} \\
\omega_{21} & 0 
\end{array} \right) |\Phi_n \rangle 
\equiv  W ~| \Phi_n \rangle \nonumber \\
&=& \sum_{k=1,2} \langle \Phi_k |W|\Phi_n\rangle
\{ A_k ~|\Phi_k\rangle + 
\sum_{l\ne k} ~B_k^l ~|\Phi_l\rangle \} \; .
\end{eqnarray}
Here
$\langle \Phi_k|\Phi_{k }\rangle  \equiv   A_k  \ge 1 $ according to 
(\ref{eif2}) and $
\langle \Phi_k|\Phi_{l\ne k }\rangle = - 
\langle \Phi_{l \ne k  }|\Phi_{k}\rangle  \equiv 
B_k^{l}, ~|B_k^{l}|\ge 0 
$ according to  (\ref{eif3}). The $A_k$ and $B_k^l$
characterize the degree of resonance overlapping.
In the regime of overlapping resonances, $1>A_k >0$, $|B_k^l| >0$, and 
equation (\ref{mix1}) is nonlinear. 
The most important part of the nonlinear contributions is contained in 
\begin{eqnarray}
\label{mix2}
(H_0  - \epsilon_n) ~| \Phi_n \rangle =
\langle \Phi_n|W|\Phi_n\rangle ~|\Phi_n|^2 ~|\Phi_n\rangle  
\end{eqnarray}
which is a nonlinear Schr\"odinger equation. According to
(\ref{mix1}), the nonlinear Schr\"odinger equation (\ref{mix2})
passes smoothly into the standard linear Schr\"odinger equation
when $A_k \to 1$,  $B_k^l \to 0$ and $r_k \to 1$, i.e. far from an EP. 

The cross section can be calculated from the $S$ matrix, $\sigma (E)
\propto |1-S(E)|^2$. A unitary representation of the $S$ matrix in the
case of two resonance states coupled to one common continuum of
scattering wavefunctions reads \cite{ro03}
\begin{eqnarray}
\label{smatr}
S = \frac{(E-E_1+\frac{i}{2}\Gamma_1)~(E-E_2+\frac{i}{2}\Gamma_2)}{(E-E_1-
\frac{i}{2}\Gamma_1)~(E-E_2-\frac{i}{2}\Gamma_2)}
\end{eqnarray}
where $E_i$ and $\Gamma_i$ are defined in (\ref{eiv1}) and $E$ is the
energy of the system. This representation of the $S$ matrix contains
the influence of EPs onto the cross section via the eigenvalues of 
$\ch^{(2)}$, Eq. (\ref{eiv1}). It provides reliable results therefore 
also when $r_k < 1$ ~\cite{ro03}.

At a double pole of the $S$ matrix (being an EP), 
the resonance line shape deviates
from the Breit-Wigner one. In this case, the $S$ matrix reads \cite{ro03}
\begin{eqnarray}
\label{smatr2}
S = 1-2i\frac{\Gamma_d}{E-E_d-\frac{i}{2}\Gamma_d}-
\frac{\Gamma_d^2}{(E-E_d-\frac{i}{2}\Gamma_d)^2}
\end{eqnarray}
where $E_1=E_2\equiv E_d$ and $\Gamma_1=\Gamma_2\equiv \Gamma_d$.
The second term corresponds to the usual linear term (however with the
factor $2$ in front) while the third term is quadratic. The
interference between these two parts of the $S$ matrix has been
illustrated by, e.g., Fig. 9 in
\cite{mudiisro} where the cross section is calculated
for the case of two resonance states coupled to one decay channel. The
asymmetry of the line shape of both peaks in the cross section 
at the double pole of the $S$ matrix is described by (\ref{smatr2}).

\section{Eigenvalues and eigenfunctions of the non-Hermitian operator 
in the case with $N > 2$ states coupled to one common continuum}
\label{more}

Knowing the properties of the eigenvalues and eigenfunctions of ${\cal
  H}^{(2)}$, Eq.
(\ref{int3}), in the neighborhood of an EP, it is interesting to study
the more general case with  $N>2$ states which are  coupled to one
common continuum of scattering wavefunctions. For this purpose, we 
consider an $N\times N$ matrix 
\begin{eqnarray}
\hspace*{-2cm}
{\cal H}^{(N)} = 
\left( \begin{array}{cccc}
\epsilon_{1} = \varepsilon_{1}+\omega_{11}~~~  & 0 & \ldots &\omega_{1N}   \\
0 & ~~~\epsilon_{2} = \varepsilon_{2}+\omega_{22}~~~  &  \ldots & \omega_{2N}\\
\vdots     & \vdots &             \ddots&   \vdots \\
\omega_{N1} & \omega_{N2}       &    \ldots   & ~~~\epsilon_{N} =
 \varepsilon_{N}+\omega_{NN} \\
\end{array} \right) 
\label{form1}
\end{eqnarray}
the diagonal elements of which are the $N$ complex eigenvalues 
$ \varepsilon_{i} + \omega_{ii}
\equiv e_i + i/2~\gamma_i$ of a non-Hermitian operator \cite{comment2}. 
The $\omega_{ii}$ are the so-called selfenergies of the states
arising from their coupling  to the environment of scattering
wavefunctions into which the system is embedded.
In atomic physics, these values are  known as Lamb shift.
Our calculations are performed with  coupling matrix elements
$\omega_{ii}$ the values of which do not depend on the parameter considered. 
In such a case, the $\omega_{ii}$ can 
considered   to be included into the diagonal matrix elements,
which read  $\epsilon_i \equiv \varepsilon_{i}+\omega_{ii} =  
e_i + i/2~\gamma_i$. The $e_i$ and  $\gamma_i$ denote the 
energies and widths, respectively, of the $N$ states (including their
selfenergies) without account of the interaction 
of the different states via the environment. The
$\omega_{kN} = \omega_{Nk}, ~k=i,j$ simulate the interaction of
the two states $i$ and $j\ne i$ via the common environment (consisting
of one continuum of scattering wavefunctions) \cite{simulation}, 
see Eqs. (\ref{form5}) to (\ref{form12}) in the Appendix 
\ref{ap2}.  This interaction is
important only at high level density and near to an EP.

In the Feshbach projection operator formalism (see Appendix
\ref{ap2}), the internal
(real) interaction of two states $i$ and $j\ne i$
(which appears in the closed system described by the Hermitian
Hamiltonian $H_B$) is taken into account by
diagonalizing $H_B$, see Eq. (\ref{form5}). 
The external interaction of these states
(via the environment) is contained in the $\omega_{ij}$ which is complex, 
see Eqs. (\ref{form11}) and (\ref{form12}). 
Im$(\omega_{ij})$ becomes  important at high level
density where the corresponding resonance states overlap.

In the present paper, we are interested in the case that all $N$ states
are coupled to one another
via a common continuum of scattering wavefunctions. 
Such a case can be represented by (\ref{form1}), i.e. by
assuming $\omega_{i~k\ne N} = \omega_{k\ne N~i} = 0$ 
while  $\omega_{iN} = \omega_{Ni} \ne
0$ \cite{simulation}. The results of some numerical
studies will be given in Sect. \ref{num2}. They show 
width bifurcation under the influence of EPs also in a system with $N>2$ states.

We add here some remarks on the  symmetry  in the 
neighborhood of an EP. An EP that is well separated from the influence of
external sources (including the influence caused by
other resonance states), is highly symmetric when
approached by varying  a certain parameter. 
That means, the two states pass one into the other one
according to (\ref{eif5}) with an exchange of their wavefunctions.
At a certain finite distance from the
EP, there are again two states with the wavefunctions 
$|\Phi_1|$ and  $|\Phi_2|$, respectively.

In the neighborhood of an EP, symmetry violation appears under the 
influence of another resonance state 
due to the finite parameter range around the EP in which the 
wavefunctions of the two states are mixed  with each other \cite{ro01}. 
When the interaction of the third state is symmetric relative to
the two crossing ones, the third state will appear as an {\it
observer}, i.e. it will not contribute actively to the spectroscopic  
redistribution processes caused by the EP.
Numerical examples of such a situation are shown in the transmission 
through a quantum dot \cite{trsbsadreev} and also in the generic case
studied in \cite{fortsch2,elro2}.
When the interaction of the third state  with the two
crossing ones is, however, not symmetrically,  irreversible
processes may appear
due to the nonlinear terms in the Schr\"odinger equation
(\ref{mix2}). Numerical examples will be shown in Sect.
\ref{num2}.

\section{Numerical results for the eigenvalues and eigenfunctions of
$\ch$ with $N=2$ states}
\label{num1}

The calculations are performed with the Hamiltonian $\ch^{(2)}$, 
Eq. (\ref{int3}). The $\epsilon_i = \epsilon_i(a) =
e_i(a)+\frac{i}{2}\gamma_i$ are the complex energies of two
states, including their self-energies. The energies $e_i(a)$ 
depend on a certain parameter $a$ while the $\gamma_i$ are 
fixed and constant in the parameter range considered.
The $\omega_{ij}=\omega_{ij}(a)$  stand for
the external interaction of the two states via the environment. 
According to the results \cite{ro01} 
for exact calculations with real $\omega$, the wavefunctions of the two states 
are mixed in a finite parameter range around the critical value of 
their crossing. We simulate this fact 
by assuming a Gaussian distribution for the coupling coefficients,
\begin{eqnarray}
\omega_{ik}(a)= \omega_{ki}(a) = 
\omega\cdot exp~[-(\epsilon_i(a) - \epsilon_k(a))^2] \; .
\label{omdef}
\end {eqnarray}
The coupling coefficients $\omega$ are complex, 
generally (see Sect. \ref{eigen}).

\begin{figure}[ht]
\begin{center}
\includegraphics[width=13cm,height=8cm]{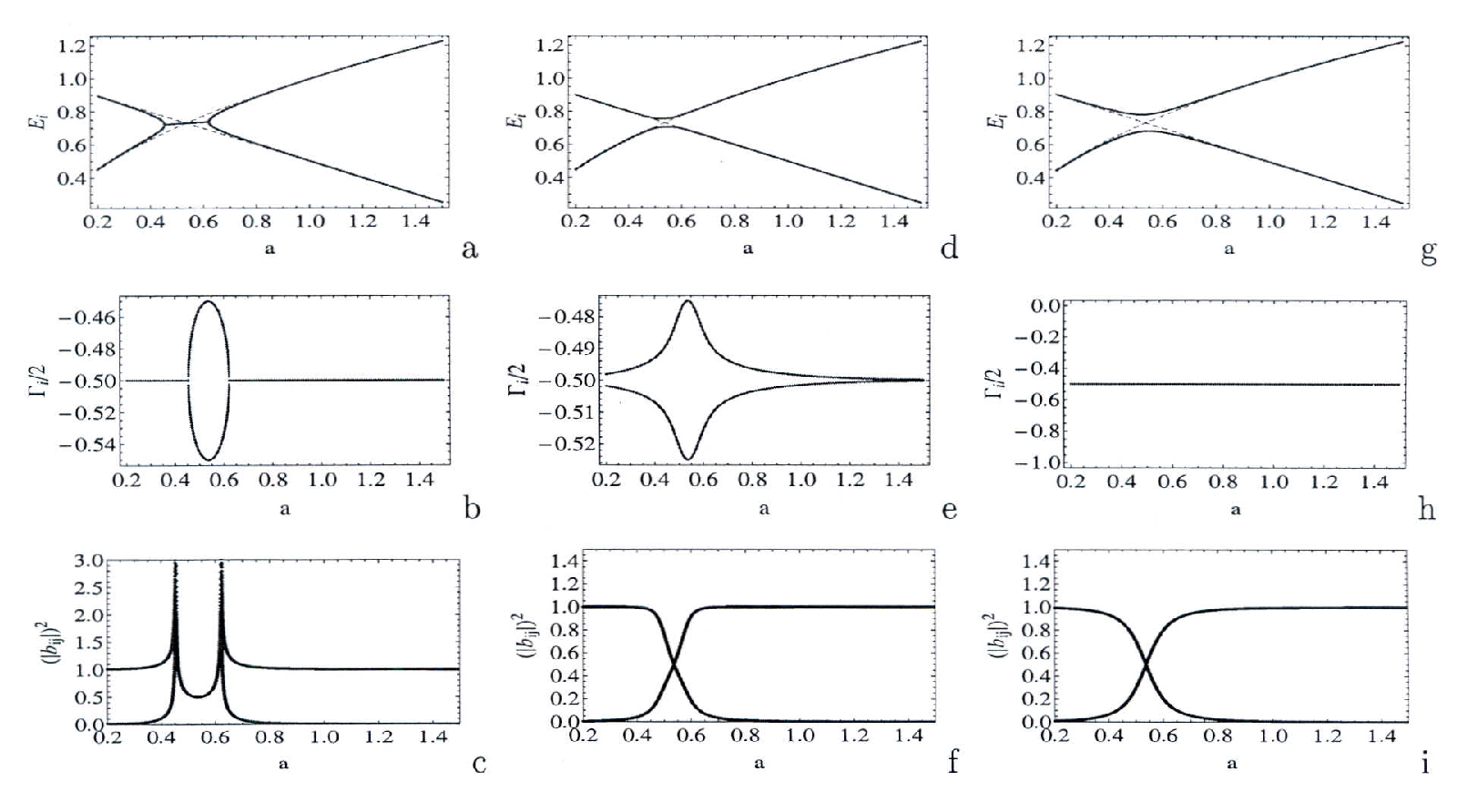}
\end{center}
\caption{\footnotesize
Energies $E_i$ (full lines), widths $\Gamma_i/2$ 
and mixing coefficients $|b_{ij}|^2$ of $N=2$
states coupled to $K=1$ channel as a function of the parameter $a$. 
The parameters of the subfigures are
$\omega=0.05~i$ (left panel), $\omega=0.025~(1+i)$ (middle panel) and
$\omega=0.05$ (right panel). Further parameters:
$\gamma_1/2 = \gamma_2/2 =- 0.5, ~e_1=1-a/2; ~e_2=\sqrt{a}$.
The dashed lines show $e_i(a)$. \\
}
\label{fig1}
\end{figure}

\begin{figure}[ht]
\begin{center}
\includegraphics[width=13cm,height=8cm]{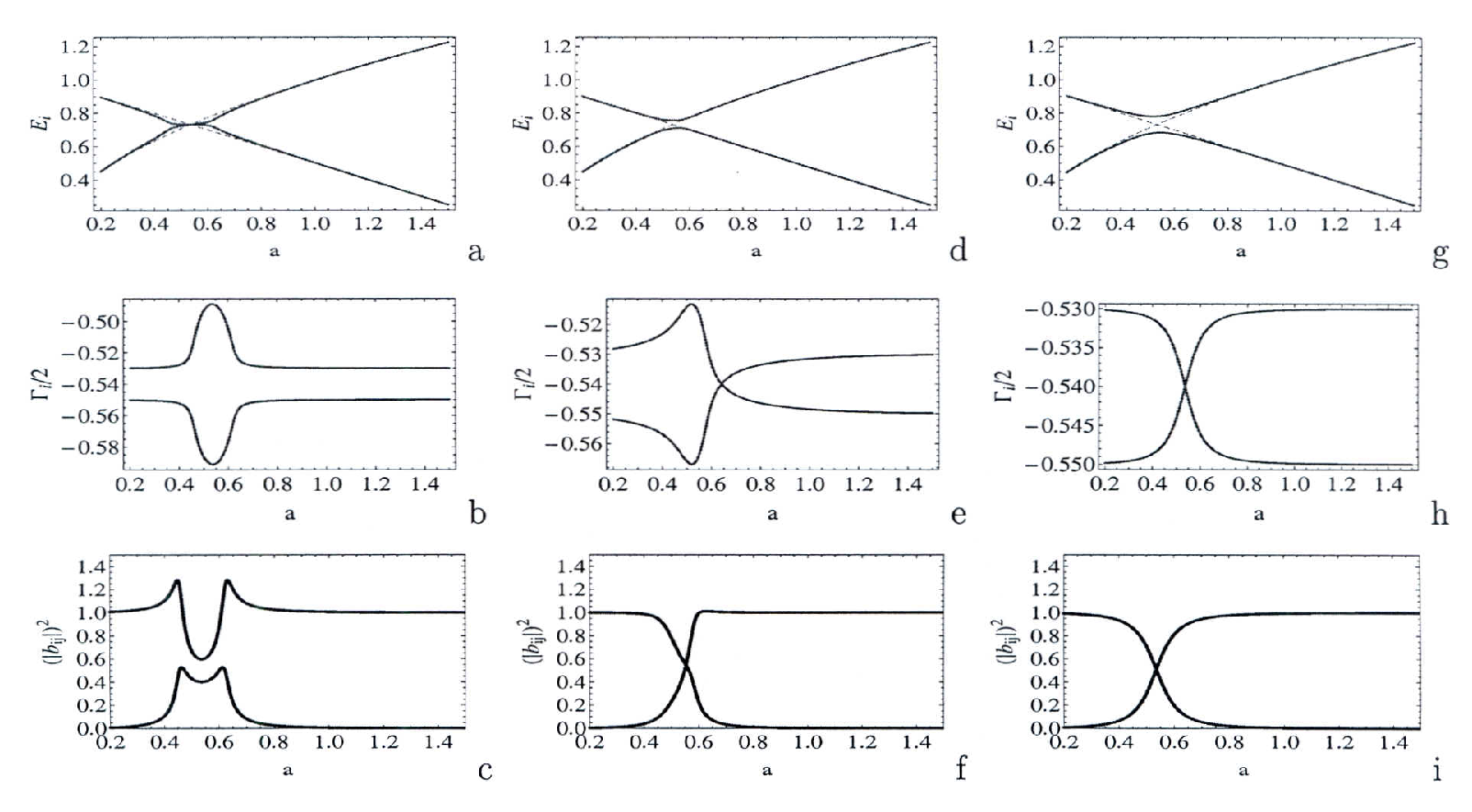}
\end{center}
\caption{\footnotesize
The same as Fig. \ref{fig1}, but $\gamma_1/2 =- 0.53; ~\gamma_2/2 =- 0.55$.\\
}
\label{fig2}
\end{figure}

We are interested, above all, in the situation at high level density
where the resonance states overlap and, according to Sect. \ref{eigen},
the influence of EPs 
onto the eigenvalues and eigenfunctions of $\ch^{(2)}$ can be seen.
In Figs. \ref{fig1} and \ref{fig2}, we show results for the
eigenvalues $\ce_i = E_i +\frac{i}{2} \Gamma_i$ and the mixing
coefficients $|b_{12}|^2$, defined in (\ref{mixb}),
of two states $i=1,~2$ for real, complex as well as for
imaginary $\omega$.

\begin{figure}[ht]
\begin{center}
\includegraphics[width=13cm,height=8cm]{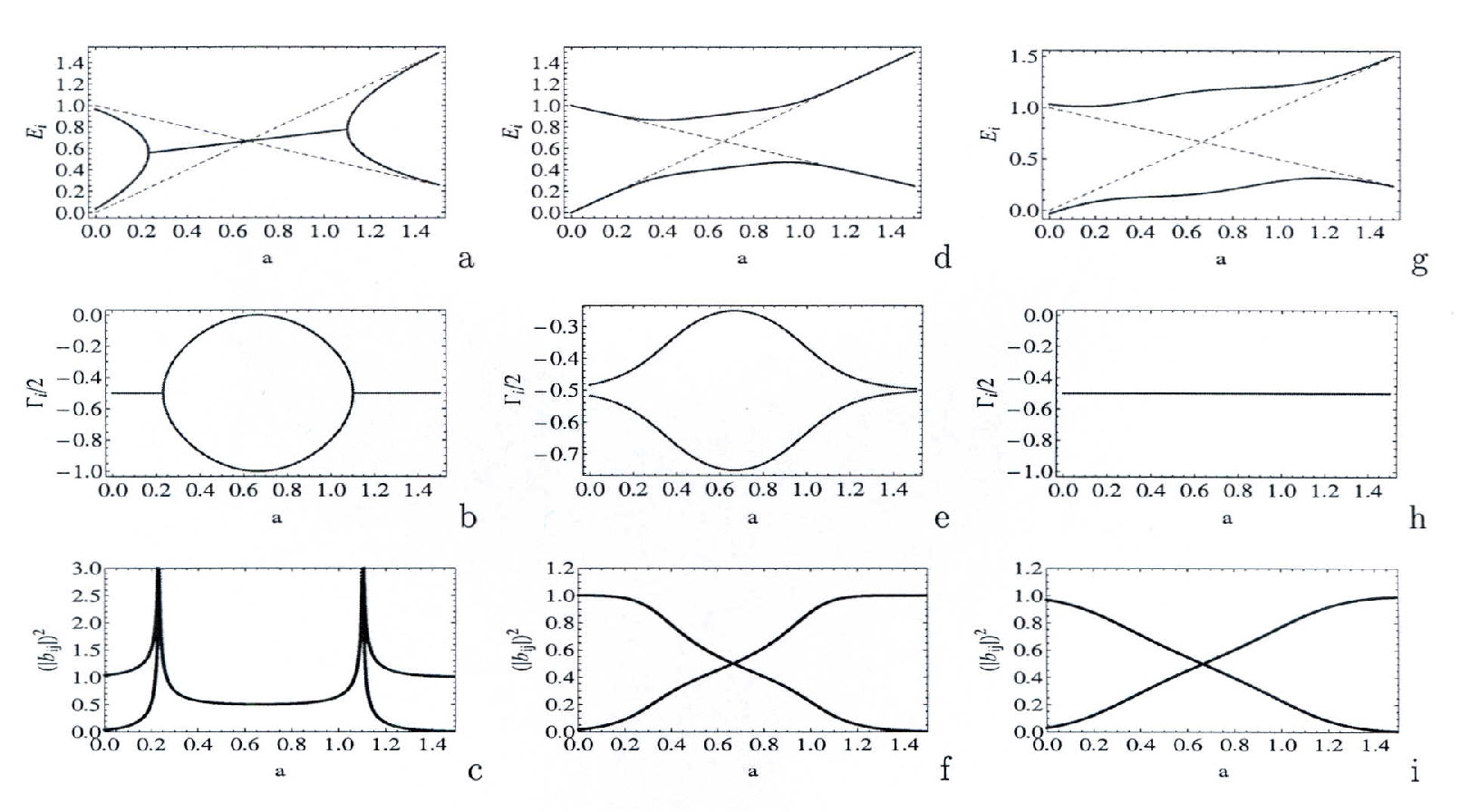}
\end{center}
\caption{\footnotesize
The same as Fig. \ref{fig1}, but $\omega = 0.5~i$ (left
panel),  ~$\omega = 0.25~(1+i)$ (middle panel),  $\omega = 0.5$
(right panel), and ~$e_2 = a$.\\
}
\label{fig3}
\end{figure}

In Fig. \ref{fig1}, the widths of both states are equal, $\gamma_1 =
\gamma_2 = -0.5$. This case  described by (\ref{int6a}) to
(\ref{int6d}), is well reproduced in the numerical simulations.
The case with real $\omega$ (right panel) shows the familiar picture
of level repulsion and an exchange of the two states at the 
critical value $a_{\rm cr}$  of the parameter $a$. The critical value  
$a_{\rm cr}$ is determined by the crossing point
of the two trajectories $e_i(a)$. The EP is beyond the parameter range
shown in the figure as discussed below (\ref{int6a}). 

The picture with imaginary  $\omega$ (left panel) is completely
different from that with real $\omega$, 
in full agreement with Eqs. (\ref{int6b}) to (\ref{int6d}).
The two EPs can be seen very clearly in the mixing coefficients
$|b_{ij}|^2$ which increase limitless in approaching the EPs (see
Sect. \ref{eigen}). According to (\ref{int6c}) and (\ref{int6d}), the
energies $E_i$ of the two states are equal in the parameter range 
between the two EPs, and level repulsion appears only beyond the
EPs. The widths $\Gamma_i$ show the opposite behavior: they are equal
beyond the EPs but bifurcate in the parameter range between the two
EPs.  Here, small variations of the parameter $a$ cause large changes
in the widths, especially near to the two EPs. It is remarkable that
the width bifurcation increases in this region
{\it without any change of the coupling strength $\omega$}. The only value
which is changed in the calculations is  the (external) parameter $a$.

The width bifurcation appears also when the coupling coefficients $\omega$
are complex (middle panel). However, the EPs can not be seen in the
parameter range shown (as in the case with real $\omega$, right panel). 
In all cases, the wavefunctions are mixed
completely  at the critical value  $a_{\rm cr}$ that is
determined by the crossing point of the two trajectories $e_i(a)$.
Only when $\omega$ is imaginary (left panel), the complete mixing occurs in a
finite range of $a$. 

Fig. \ref{fig2} shows that the main results obtained for $\gamma_1 =
\gamma_2 =- 0.5$ (Fig. \ref{fig1}) remain also when the two 
$\gamma_i$ differ from one another ($\gamma_1=-0.53, ~\gamma_2=-0.55$). 
For real $\omega$ (right panel), the energy trajectories avoid
crossing and the states are exchanged at $a=a_{\rm cr}$. However, 
the widths cross freely at the critical
point. This fact is very well known from many studies performed during
last years.  

When $\omega$ is imaginary (left panel), the energies of the two
states are equal to one another in a certain  finite parameter range.
Here, the widths bifurcate although the coupling strength $\omega$ is
fixed. Further, the wavefunctions of the two states are mixed in the 
regime with width bifurcation. Although the EPs can not be seen
in the figure, the mixing coefficients $|b_{ij}|^2$ point to their
existence not far from the considered parameter range (they can be
found by means of varying a second parameter). However, the two states 
are not exchanged.

The energy trajectories avoid crossing and the widths bifurcate 
in a finite parameter range
around $a=a_{\rm cr}$  also in the case when $\omega$ is complex (middle panel). 
In this case, the states are exchanged in a similar manner as in the
case with real $\omega$ (right panel).

The width bifurcation increases with increasing coupling strength
$\omega$ between system and environment. In Fig. \ref{fig3}, 
results are shown that are obtained for a value of $\omega$ which is
larger by a factor of ten than that used for the calculations in
Fig. \ref{fig1}. The distance between the two EPs is larger in this case 
(Fig. \ref{fig3} left panel) than in the case   with smaller
$\omega$ (Fig. \ref{fig1} left panel), what is in agreement with
(\ref{int6d}).  Furthermore, the coupling to the continuum causes a much
stronger width bifurcation (Fig. \ref{fig3}.b) than in the foregoing case
(Fig. \ref{fig1}.b). The width of one state increases at the cost 
of the width of the other state. Finally, the width of one  
state approaches zero and that of the other state $-2$. 
The difference between $\Gamma_1/2$ and $\Gamma_2/2$ is about 
$\sum_{i=1}^2 |\gamma_i|/2 = 2\cdot 0.5 = 1$. By further increasing of 
$\omega$, the system consisting of two {\it decaying states} 
will crash. 

The results for complex and real coupling strengths 
are shown in Fig. \ref{fig3},
middle and right panels, respectively.
In any case, the eigenvalues and eigenfunctions of
$\ch^{(2)}$ are influenced by the two EPs in a larger parameter range
than in Fig. \ref{fig1}.

It should be underlined here that far from the critical parameter
range shown in Figs. \ref{fig1}--\ref{fig3}, the ${\cal E}_i$
trajectories approach the $\epsilon_i$ trajectories. Here, the states
are exchanged, at most, and the wavefunctions 
are normalized in the standard manner.

\section{Numerical results for the eigenvalues and eigenfunctions of
$\ch$ with $N>2$ states}
\label{num2}

In this section, we show results obtained for the eigenvalues 
$\ce_i = E_i +\frac{i}{2} \Gamma_i$ of  (\ref{form1})
and for the mixing
coefficients $|b_{12}|^2$, defined in (\ref{mixb}) with $N>2$,
for real, complex as well as for imaginary coupling coefficients. 
As in Sect. \ref{num1} for $N=2$ states, we consider $\epsilon_i =
\epsilon_i(a) = e_i(a) +\frac{i}{2}\gamma_i$ where the energies $e_i$
depend on a certain parameter $a$ and the $\gamma_i$ are constant in
the considered parameter range. 
The fact that the wavefunctions of the states are mixed in a finite
parameter range around their crossing \cite{ro01} is simulated, in analogy
to Sect. \ref{num1}, by assuming the Gaussian distribution 
(\ref{omdef})
for the nonvanishing coupling coefficients $\omega_{ik}$ with $i=N$ or $k=N$.

\begin{figure}[ht]
\begin{center}
\includegraphics[width=13cm,height=8cm]{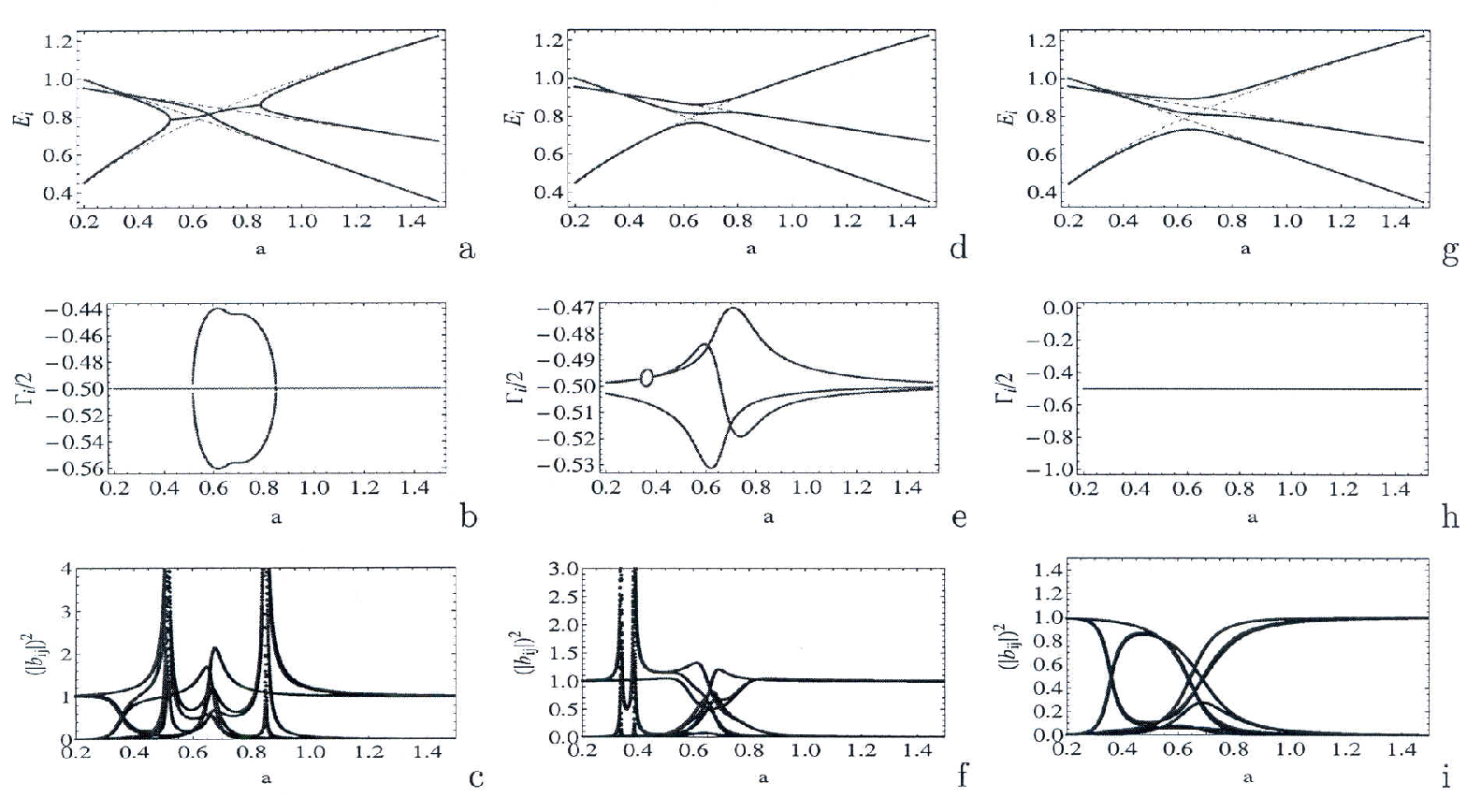}
\end{center}
\caption{\footnotesize
Energies $E_i$ (full lines), widths $\Gamma_i/2$ 
and mixing coefficients $|b_{ij}|^2$ of $N=3$
states coupled to $K=1$ channel as a function of the parameter $a$. 
The parameters of the subfigures are
$\omega=0.05~i$ (left panel); $\omega=0.025~(1+i)$ (middle panel) and
$\omega=0.05$ (right). Further parameters:
$\gamma_1/2 = \gamma_2/2 = \gamma_3/2 =- 0.5, ~e_1=1-a/4.5; ~e_2=1.1-a/2;
 ~e_3=\sqrt{a}$. The dashed lines show $e_i(a)$. 
}
\label{fig4}
\end{figure}

\begin{figure}[ht]
\begin{center}
\includegraphics[width=13cm,height=8cm]{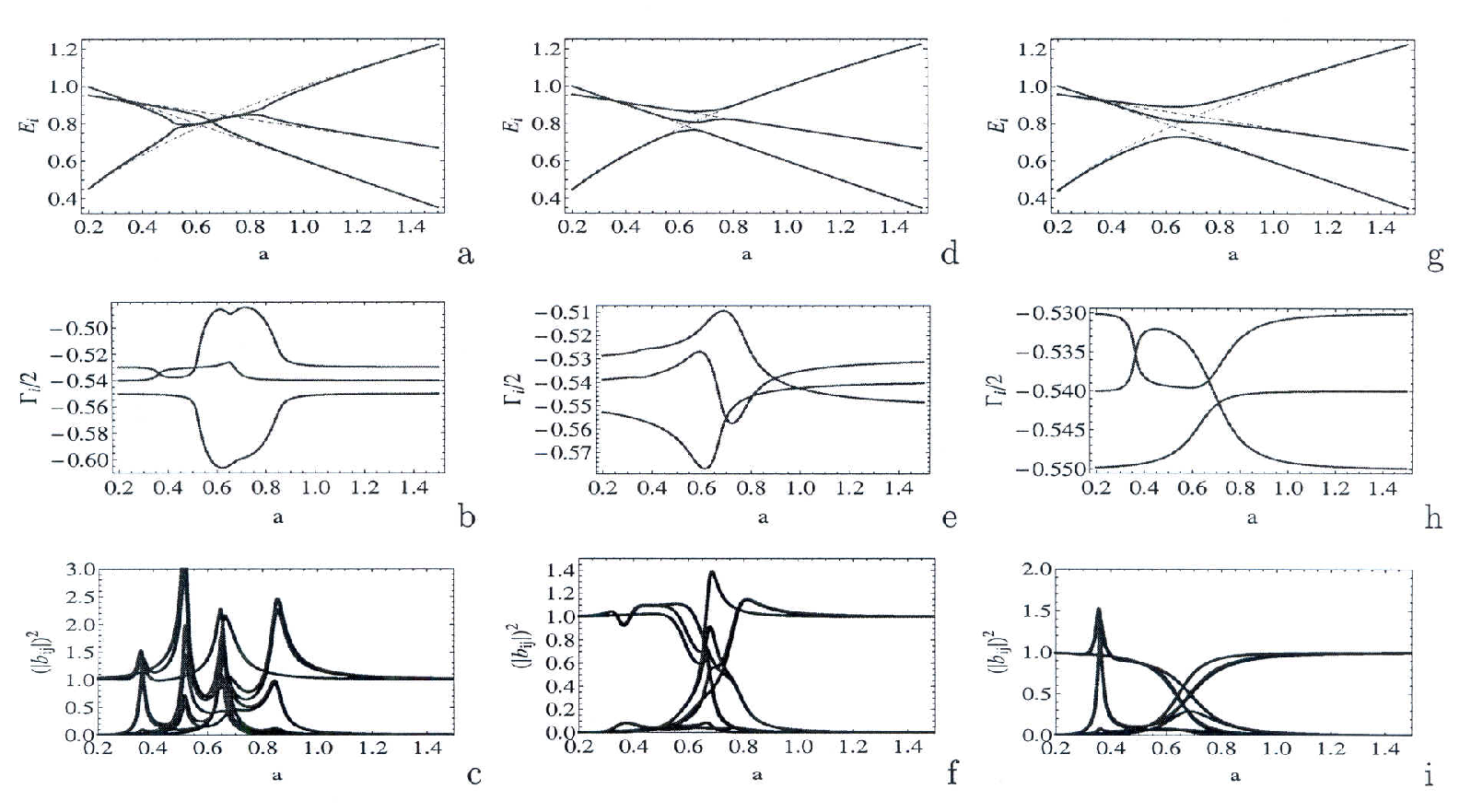}
\end{center}
\caption{\footnotesize
The same as Fig. \ref{fig4}, but 
 $\gamma_1/2=-0.53; ~\gamma_2/2=-0.54; ~\gamma_3/2 = -0.55$\\
}
\label{fig5}
\end{figure}

In Figs. \ref{fig4} and \ref{fig5}, we show some results for $N=3$.
The results are more complicated than in the case with only two crossing
states. They can be understood, however, fully  on the basis of the
results discussed in Sect. \ref{num1}. 

Fig. \ref{fig4} shows results with the same widths of all states,
$\gamma_1 = \gamma_2=\gamma_3=-0.5$.
When $\omega$ is real (right panel) 
the eigenvalue trajectories avoid crossing at different critical 
parameter values. However,
far from the critical parameter range with crossing eigenvalue
trajectories, two states are exchanged while the middle one remains
in the middle. Its trajectory is somewhat influenced in the critical region
by those of the two neighboring states. 
This result differs from those shown in \cite{fortsch2} with $N>2$ where 
the parameter dependence of all energy trajectories is the same.
In \cite{fortsch2}, the trajectory of the third (middle) state 
is not at all influenced by those of the neighboring states due 
to the high symmetry around  all the EPs in this case. 
The third state is, therefore, not involved in the
redistribution processes in the critical region and appears as an {\it
observer} state. When however the parameter dependence of the
different energy trajectories is {\it not} the same (as in Fig. \ref{fig4}),
the wavefunctions of the three
states are mixed in the  critical parameter range, see
Fig. \ref{fig4}.i  with real $\omega$.
That means, the third state loses its role as an
observer state in relation to the energy trajectories 
when the symmetry around the EPs is disturbed under the
influence of neighboring states. 

The figures with imaginary $\omega$ (left panel of Fig. \ref{fig4})
can also be understood by using the results shown in
Sect. \ref{num1}. Also in the case with $N=3$ states, two EPs can be
seen. The parameter range between them is
crossed by the  energy trajectory of the third state,
and altogether two states are exchanged.
The widths bifurcate in the range between the two EPs, however the
width of one of the states remains  constant in this range. This
behavior is similar to that studied in  \cite{fortsch2} 
with $N>2$ in which 
the parameter dependence of all energy trajectories is the same.
That means, also in the present calculation
the third state does not lose its role as an
observer state in relation to the width trajectories due to the fact that
the widths of all states are equal to one another. 
Further, the mixing coefficients $|b_{ij}|^2$ (Fig. \ref{fig4}.c)
point to another EP between the two well expressed ones at the corners
where width bifurcation starts and ends. This result can be understood
in the following manner. When the distance between the two EPs at the
corners becomes larger,  the parameter range with width bifurcation
splits into two regions.  Finally, there are two separated regions
with width bifurcation with altogether four EPs at the four corners.

Most important result with imaginary $\omega$ (left panel of 
Fig. \ref{fig4}) is that the amount of
width bifurcation depends strongly on the parameter $a$, especially in
the very neighborhood of the two EPs. Width bifurcation increases  
between the two EPs without any enhancement of the coupling strength $\omega$.  
This result  discussed in Sect. \ref{num1} for $N=2$ states, holds
also when the number of states is larger than $2$. 

Additionally, we show the results with complex $\omega$ in
Fig. \ref{fig4}, middle panel. The energy trajectories are similar to
those obtained with  real $\omega$. The width trajectories show
width bifurcation, however with a shift of the parameter values 
for the maximal and minimal value of $\Gamma$ relative to one
another. Interesting is
the appearance of an additional region with width bifurcation 
at the parameter value where two levels avoid crossing. The
corresponding two EPs can be seen in the   $|b_{ij}|^2$
(Fig. \ref{fig4}.f).

\begin{figure}[ht]
\begin{center}
\includegraphics[width=13cm,height=8cm]{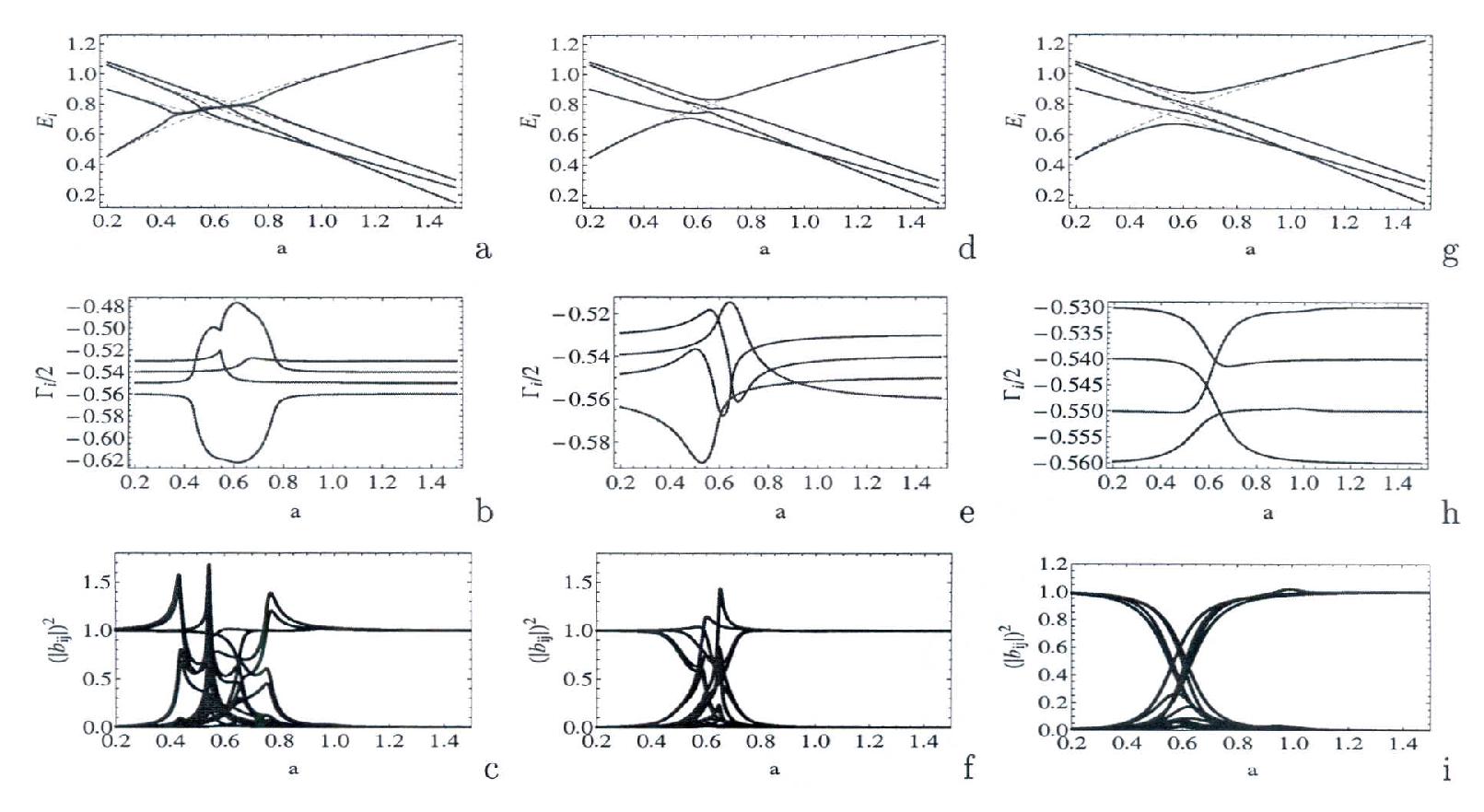}
\end{center}
\caption{\footnotesize
Energies $E_i$ (full lines), widths $\Gamma_i/2$ 
and mixing coefficients $|b_{ij}|^2$ of $N=4$
states coupled to $K=1$ channel as a function of the parameter $a$. 
The parameters of the subfigures are
$\omega=0.05~i$ (a,b,c); $\omega=0.025~(1+i)$ (d,e,f) and
$\omega=0.05$ (g,h,i). Further parameters:
$\gamma_1/2 =- 0.53; ~\gamma_2/2 =-0.54; ~\gamma_3/2 = -0.55; 
~\gamma_4/2=-0.56;
~e_1=1.2-0.7a; ~e_2=1.2-0.6a;
 ~e_3=1-0.5a; ~e_4=\sqrt{a}$. The dashed lines show $e_i(a)$. \\
}
\label{fig6}
\end{figure}

The results obtained for $N=3$ states with different widths are shown in
Fig. \ref{fig5}. Here $\gamma_1=-0.53, ~\gamma_2=-0.54, ~\gamma_3=-0.55$. 
In difference to the results shown in
Fig. \ref{fig4}, the symmetry around the EPs in relation to the
widths is eliminated in these calculations. 

For real $\omega$ (right panel of Fig. \ref{fig5}), 
we see level repulsion similar to
that in the corresponding Fig. \ref{fig4} with equal widths $\gamma_i$.
The $\Gamma_i$ trajectories show free crossings at parameter values
where the energy trajectories $E_i$ avoid crossing. The results point
to a new EP at a small $a$ value where both $E_i$ and $\Gamma_i$ of
both states are nearly the same. 

Also the results for imaginary $\omega$ (left panel of Fig. \ref{fig5})
point to the new EP at a small $a$
value. Most important is however that width bifurcation appears in 
almost the same parameter range as in the corresponding Fig. \ref{fig4}.b
with equal $\gamma_i$. Also in Fig. \ref{fig5}.b, width bifurcation
changes by varying the parameter $a$ (especially in the very
neighborhood of the EPs) by keeping fixed the coupling strength
$\omega$. 

The results for complex $\omega$ show level repulsion and 
width bifurcation. However also
in this case  the maximal and minimal values of the widths are at slightly 
different parameter values $a$. 

Neither for imaginary nor for complex or real $\omega$, an observer
state is observed in Fig. \ref{fig5}. This is due to the fact that 
the symmetry around the EPs is disturbed in these calculations not 
only in relation to the energy trajectories
but also in relation to the width trajectories.  
Such a situation is, of course, more realistic than that with highly
symmetrical energy and width trajectories. 

The results of calculations with $N>3$ states and different widths
$\gamma_i$  show all the
characteristic features discussed above for $N=3$ states, see
e.g. Fig. \ref{fig6}. Most important result is that width bifurcation 
occurs in the very neighborhood
of EPs  and remains in a comparably large parameter range when
$\omega$ is imaginary (Fig. \ref{fig6} left panel). 
Here, the wavefunctions are strongly mixed.
The eigenvalue trajectories obtained with real  $\omega$ 
(Fig. \ref{fig6} right panel) are completely different 
from those with imaginary $\omega$ (Fig. \ref{fig6} left panel).

\section{Influence of exceptional points onto the cross section with 
$N=2$ states}
\label{num3}

We consider now the influence of EPs onto the cross section of a
reaction with $N=2$ resonance states. We are interested, above all, in
the transition from the case in which the two resonance states are 
well isolated from one another to the case where they are strongly 
overlapping. In the
first case, the coupling of the two states via the continuum of scattering
states is (almost) negligible. According to the results obtained
and discussed in
the foregoing sections, it plays, however, an important role in the
second case because the eigenvalues and eigenfunctions of $\ch^{(2)}$
are strongly influenced by EPs at high level density.

The cross section is calculated
from $\sigma (E) \propto |1-S(E)|^2$ with the $S$ matrix (\ref{smatr})
which contains the eigenvalues of $\ch^{(2)}$.
The eigenvalues and eigenfunctions of $\ch^{(2)}$
are calculated as a function of a certain parameter $a$ 
in a similar manner as in Sect. \ref{num1}. The coupling matrix elements are 
\begin{eqnarray}
\hspace*{-1.8cm}
\omega_{12}(y)=\omega_{21}(y) =
\tilde\omega\cdot \bigg(\sqrt{1-y^2} + iy\bigg); ~~~~ \tilde\omega=\omega_0\cdot
exp[-(\epsilon_2(a) - \epsilon_1(a))^2] \; .
\label{om0def}
\end {eqnarray}
Here the ratio between the real and imaginary parts is explictly
expressed by the factor $\sqrt{1-y^2} + iy$. The limiting cases $y=0$ and
$y=1$ correspond to real and imaginary coupling coefficients, respectively.
The $\tilde\omega$ depend on the parameter $a$ and the $\omega_0$ are
real numbers that characterize merely the coupling strength. 
We mention here, that the assumption (\ref{om0def}) for
the $\tilde\omega$ is not decisive for the results
obtained. Qualitatively the same results are obtained with, e.g., a
linear relation between the real and imaginary parts of $\tilde\omega$.

\begin{figure}[ht]
\begin{center}
\includegraphics[width=12cm,height=12cm ]{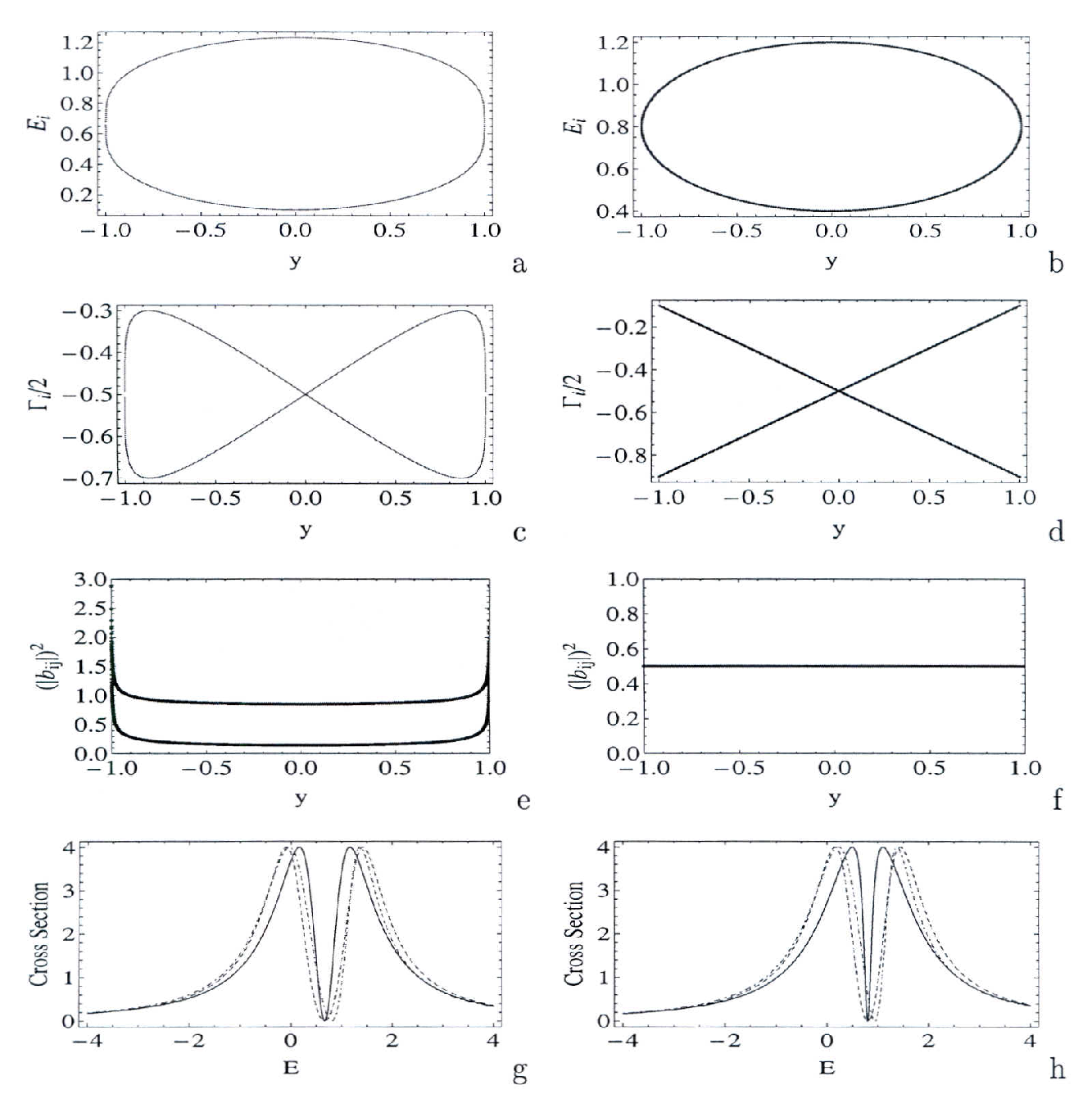}
\end{center}
\caption{\footnotesize
Energies $E_i$, widths $\Gamma_i/2$ and 
mixing coefficients $|b_{ij}|^2$ of $N=2$
states coupled to $K=1$ channel as a function of $y$ and cross section 
with 2 resonances as a function of $E$ for the small value  $\omega_0 = 0.4$. 
The value $a=0.26666$   corresponds  to an EP ($a=a_1$, left panel) and
the value  $a=0.8$ to the maximum  width bifurcation ($a=a_0$, right panel).  
The cross section is calculated at $y=0$ (dashed), $y=0.5$ (dash-dotted) and
$y=1$ (solid).  
The further parameters  are 
$~e_1=1.2-0.5~a$; $~e_2=a;
 ~\gamma_1/2 = \gamma_2/2 =-0.5$.
}
\label{fig7}
\end{figure}

\begin{figure}[ht]
\begin{center}
\includegraphics[width=12cm,height=12cm]{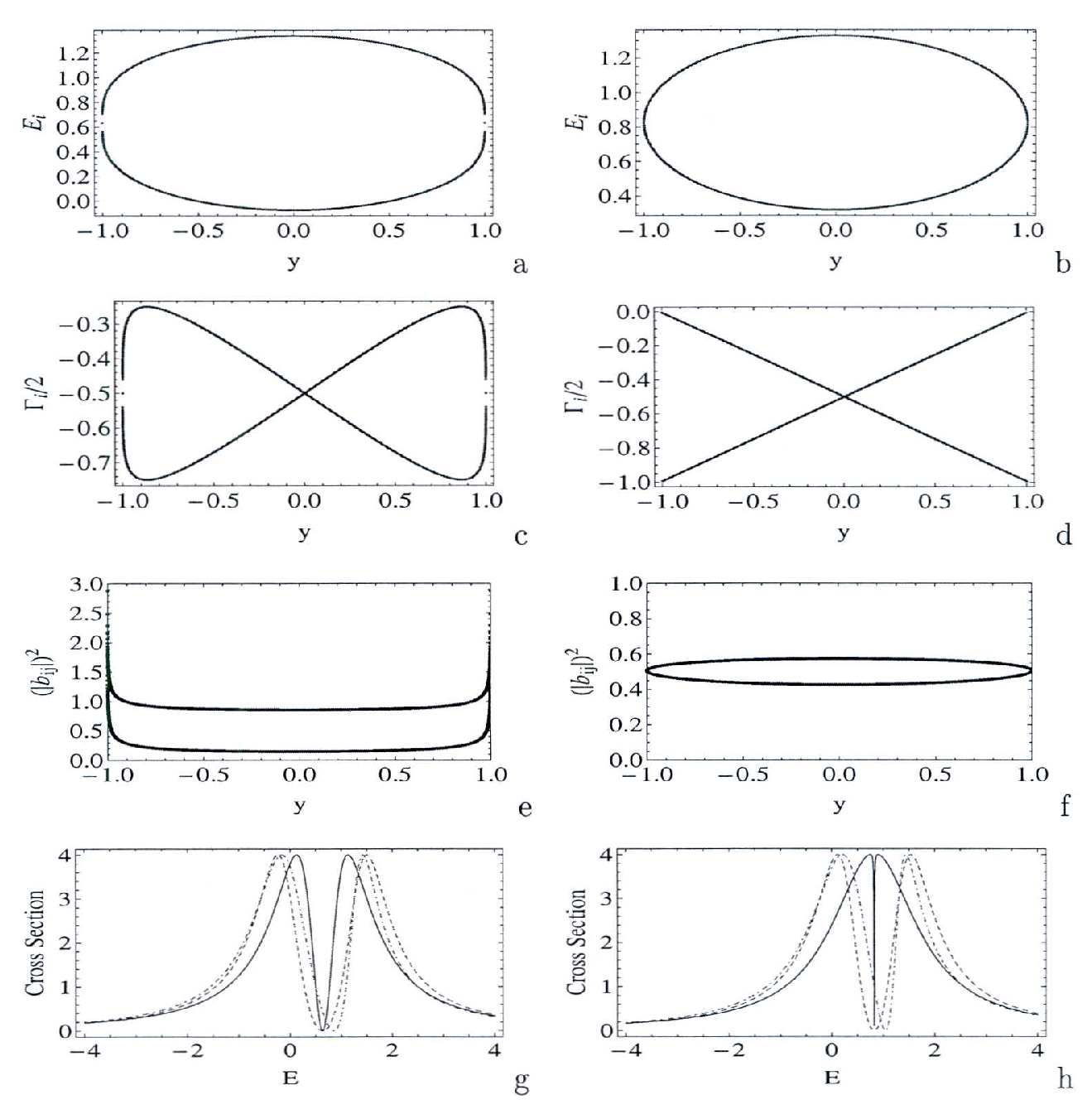}
\end{center}
\caption{\footnotesize
Energies $E_i$, widths $\Gamma_i/2$ and 
mixing coefficients $|b_{ij}|^2$  of $N=2$
states coupled to $K=1$ channel as a function of $y$ and cross section 
with 2 resonances as a function of $E$ for the  value  $\omega_0 = 0.5$. 
The value $a=0.133333$ (left panel) corresponds  to an EP ($a=a_1$)
and the value $a= 0.9$ (right panel) corresponds to  width bifurcation ($a\ne
a_0$ and $a_1 < a < ~a_2$). 
The cross section is calculated at $y=0$ (dashed), $y=0.5$ (dash-dotted) and
$y=1$ (solid).
The further parameters  are 
$~e_1=1.2-0.5~a$; $~e_2=a; ~\gamma_1/2 = \gamma_2/2 =-0.5$.
}
\label{fig8}
\end{figure}

We calculate the cross section for the case of equal 
widths of the two states, $\gamma_1 = \gamma_2$.
According to (\ref{int6b}),  (\ref{int6c}) and (\ref{int6d}), 
we have two EPs. Between the two EPs, it is $y=1$ and the widths bifurcate. 
This analytical result agrees with those of numerical studies,
see Sect. \ref{num1}.

In the following, the two EPs are denoted by EP1 and EP2 
and the corresponding values of the parameter $a$ by $a_1$ and $a_2$,
respectively. Near to EP1 and EP2, i.e. at the corners of the
parameter range $a_1\le a \le a_2$,
width bifurcation increases quickly. The parameter value with maximal
width bifurcation (in the middle between EP1 and EP2) will be denoted by $a_0$. 
As follows immediately from  (\ref{form11}) and  (\ref{form12})
as well as from (\ref{int6d}), it is $y=1$ for all parameter
values  $a_1 \le a \le a_2$ (under the condition $\gamma_1 = \gamma_2$).  
For parameter values beyond this region, $\omega_{ik}$ is complex.
For resonance states that are well separated in energy, 
$y \to 0$.

In Figs. \ref{fig7}.a-f and \ref{fig8}.a-f, we show the energies $E_i$, 
widths $\Gamma_i$ and
mixing coefficients $|b_{ij}|^2$  as a function of $y$ calculated with  
$\omega_0=0.4$ and $\omega_0=0.5$, respectively.
Similar pictures are obtained  for $\omega_0=0.1, ~0.3, ~0.6$ and 0.7.
When $a=a_1$ or $a=a_2$, the characteristic features of an EP can be
seen at $y=1$: $E_1\to E_2, ~\Gamma_1\to \Gamma_2$ and $b_{ij} \to
\infty$ (Figs. \ref{fig7}.a,c,e and \ref{fig8}.a,c,e). 
When $a_1< a < a_2$, the EPs do not show up  in the figures (Figs. 
\ref{fig7}.b,d,f and \ref{fig8}.b,d,f). However, 
the wavefunctions $\Phi_i$ of both states are, in this case, strongly 
mixed in the basic wavefunctions $\Phi_j^0$
(Figs. \ref{fig7}.f and \ref{fig8}.f).
This result corresponds to those shown in Sect. \ref{num1}.

In Figs. \ref{fig7} and \ref{fig8}, also the
cross section  $\sigma (E)$ for three different values of the
parameter $a$ is shown. 
The value $\omega_0=0.4$ of the coupling strength in Fig. \ref{fig7}
is relatively small: at $\omega_0=0.1$, the calculated cross
section does not at all depend on $a$, while a small  
variation of the cross section with different $a$ values can be seen
at $\omega_0=0.3$ (as additional calculations have shown). 
This variation  is however smaller than that in a calculation with
$\omega_0=0.4$ (Fig. \ref{fig7}). It 
increases further when $\omega_0=0.5$  (Fig. \ref{fig8}).

\begin{figure}[ht]
\begin{center}
\includegraphics[width=12cm,height=11cm]{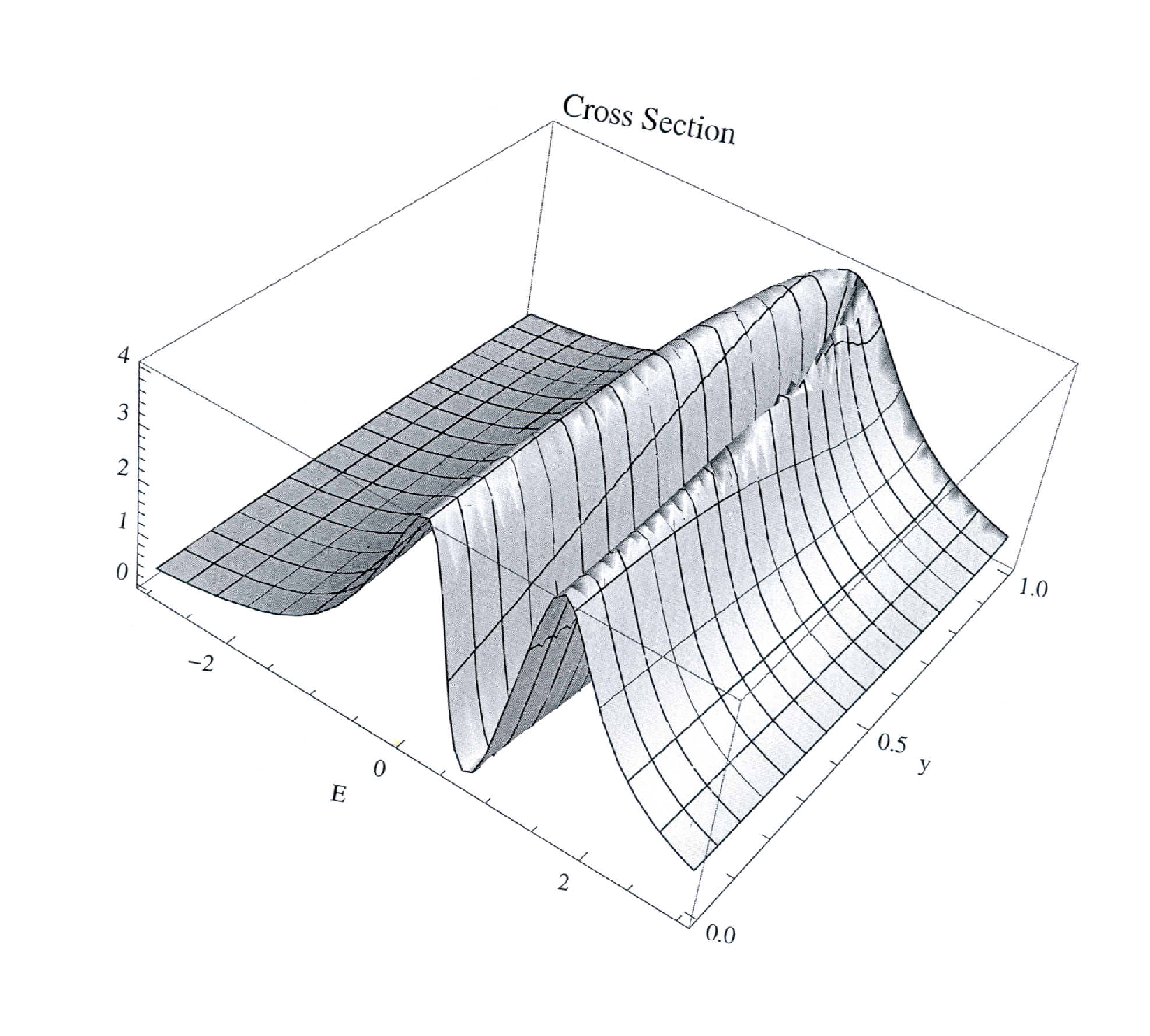} 
\end{center}
\caption{\footnotesize
Cross section with 2 resonances as a function of $y$ and 
$E$ for the  value  $\omega_0 = 0.5$. 
The further parameters  are 
$~e_1=1-a/2; ~e_2=a; ~\gamma_1/2 = \gamma_2/2 =-0.5$ (compare Fig. \ref{fig3})
and $a=0.7$.\\
}
\label{fig9}
\end{figure}

The line shape obtained for $a=a_1$ (or $a=a_2$) at $\omega_0=0.4$ and 
$\omega_0=0.5$ is shown in Fig. \ref{fig7}.g and   Fig. \ref{fig8}.g,
respectively. It is described well by Eq. (\ref{smatr2})~: The
cross section calculated at an EP
($a=a_1$ and $a=a_2$, respectively) is an 
interference picture of two resonances with a relatively broad dip 
between the two bumps.  This picture is obtained analytically as
well as numerically at a double pole of the $S$ matrix \cite{ro03,marost03}
(which is nothing but an EP, see Sect. \ref{eigen}).  

When $a_1<a<a_2$, a narrow dip appears in the cross section, 
see Fig. \ref{fig8}.h for  $\omega_0 = 0.5$. 
The dip is somewhat broader when  $\omega_0<0.5$, see 
Fig. \ref{fig7}.h for  $\omega_0=0.4$. 

Most interesting result is therefore the following.
In the case with equal widths $\gamma_i$ (and $\omega_0\approx |\gamma_i/2|$), 
the dip between the two maxima in the cross section is narrower when 
$a_1 < a < a_0$ (or  $a_2 > a > a_0$) than at $a_1$ (or $a_2$) although 
the coupling strength $\omega_0$ is {\it the same} in both cases
(fixed at a sufficiently high value).
A narrow  dip appears due to the interference of two states with very
different widths. It corresponds therefore to a large width bifurcation.

Thus, the results obtained for the cross section are  
in agreement with all the results shown and discussed
in Sect. \ref{num1} (and  also with those for $N>2$ states discussed
in Sect. \ref{num2}). The width bifurcation starts beyond the EP 
{\it without  any enhancement} of the coupling strength between system
and environment.

In Fig. \ref{fig9}, the cross section is shown as a function of $y$
and $E$ calculated  at $\omega_0=0.5$. The other parameters are similar
to those in Figs. \ref{fig7} and \ref{fig8}. At small values $y$, two
well separated resonances appear in the cross section. Here, the
resonance states do not overlap and $y \approx 0$. With increasing
degree of overlapping, $y$ increases and reaches the value $y=1$ at
the EP. The value $y=1$ remains constant for all parameter values 
$a \to a_0$. 
Here width bifurcation appears and the dip between the two maxima in
the cross section becomes narrower. 
 
It should be underlined  that the maximum height of the 
resonance peaks in the cross section 
is normalized to 4 for all values of the parameter $a$ according to 
(\ref{smatr}). An enhancement of the cross
section with $y\to 1$ does therefore not appear in Fig. \ref{fig9}.
This is in contrast to calculations for a realistic system, see    
\cite{burosa} where the cross section is calculated without the 
normalization in (\ref{smatr}).

\section{Discussion of the results}
\label{disc}

\subsection{Symmetry around exceptional points  and dynamical phase transitions}
\label{disc1}

In the present paper, the properties of EPs and their influence on the
dynamics of open quantum systems are considered. Since EPs are
singular points, analytical studies are possible only in a few special
cases, see Sect. \ref{eigen}. The results of these
analytical studies  agree well with those obtained numerically.

We simulate the main features of the dynamics
of an open quantum system by considering
the Hamiltonian matrix  (\ref{int3}) and (\ref{form1}), respectively.
The physical meaning of all matrix elements  is derived from 
the basic equations (\ref{form3}) to (\ref{form12}) that characterize
open quantum systems, i.e. quantum systems that are embedded into a
continuum of scattering wavefunctions. This  environment exists
always. It can be changed by means of some external parameters,
however never be deleted. Of special interest is the fact that the
coupling matrix elements between system and environment are complex,
generally, with the real part arising from the principal value
integral (\ref{form11}) and the imaginary part originating from the residuum 
(\ref{form12}). 
We are interested in the symmetry properties around an  EP which play
an important role for a 
dynamical phase transition (DPT).

Let us first discuss the influence of symmetries around 
EPs onto  width bifurcation in a many-level system.
In the neighborhood of an EP,
we disturb first the symmetry in energy (Fig. \ref{fig4})
and then in time (inverse proportional to the 
decay width, Fig. \ref{fig5}) by the existence of a nearby state. 
Such a situation corresponds to that characteristic of a realistic system.
The results (Figs. \ref{fig4} to \ref{fig6}, $N=3$ and 4,
respectively) show that 
{\it all} states take part in the process of width bifurcation, if
the symmetries around the EPs are disturbed. 
We conclude therefore that in a realistic system at high level density
a short-lived state appears as the result of a stepwise enhancement 
of its width at the cost of the widths of  other states which
decouple more or less from the environment. This result confirms the
assumption made earlier that a DPT occurs at high level density  
by successive resonance trapping \cite{hierarch}.
According to the present results, resonance trapping takes place, all in all,
in a certain finite parameter range where the
symmetries around the EPs are violated by nearby states
due to the high level density,  and the
phase rigidity $r_i$ of the participating states $i$ is reduced ($r_i<1$).

As to a two-level system at large coupling strength $\omega$, the
number of EPs is two, and the surrounding area of each of the two EPs 
is not symmetrical. The EPs are approached
by decreasing level repulsion on the one side
and by  increasing width bifurcation on the other
side, see Eqs. (\ref{int6c}) and (\ref{int6d}) and the corresponding 
Fig. \ref{fig3} (left panel). 
Figs. \ref{fig7} to \ref{fig9} show further that 
a DPT may arise even at $N=2$
(if the interaction between system and environment is sufficiently
strong, $\omega_0 \approx |\gamma_i/2|$). 
In such a case, the parameter range between the two EPs is large and width
bifurcation is strong (Fig. \ref{fig3}). It starts beyond the EP without
any further amplifying of the coupling strength $\omega$. Finally,
width bifurcation creates one short-lived state together with a very
narrow long-lived state. Due to interferences,  a bump with a narrow
dip can be seen in the cross section.

Thus, width bifurcation occurs in both cases in a certain finite
parameter range. The difference between the two cases is that the
short-lived state  results from  successive crossings with many
other states   in the first case 
while it results from only one crossing in the second case  
(under the condition that the interaction $\omega$ between system and 
environment is  sufficiently strong).

The relation of DPTs to observables in  real many-body systems
is studied, e.g.,  in \cite{burosa}.
Reduced phase rigidity and width bifurcation cause an enhanced transmission
through a quantum dot in the neighborhood of an EP where  
$r_i < 1$ for the phase rigidity of the participating states $i$. It  
is limited in a natural manner in a realistic system
because  the $S$ matrix describes a system with {\it decaying} states
(i.e. $\Gamma_i \le 0$ for {\it all} states $i$).
Other examples for the observation of
DPTs occurring in many-particle open quantum systems at high level density
are discussed in \cite{phaselapses2,cnr11}. 

A DPT in a two-level system is observed experimentally and related to
EPs in \cite{past1,past2}. Here, 
the oscillatory dynamics of a quantum two-level system can,
in the presence of an environment, undergo a DPT to a non-oscillatory phase.
Surely, the Dicke superradiance known since many years \cite{dicke}
is, at least partly,  also related to a DPT, see the
next section.

\subsection{Dynamical phase transitions and Dicke superradiance}
\label{disc2}

Interesting is a comparison of the results obtained in the present
paper with some results known in optics.  
Many years ago, Fano studied  two nearly resonant modes that decay via
a common channel \cite{fano}. He showed that the shared decay channel yields  
additional cross coupling between the modes by means of which the
asymmetric line shape for electron scattering from helium can be
explained. According to present-day studies, the line shape
of two neighboring resonant states with almost the same lifetimes 
is described by the $S$ matrix (\ref{smatr2}) at (or near to) 
a double pole. As shown in Sects. \ref{num1} and \ref{num3}, the 
interference picture may change to a bump with an extremely sharp dip by 
varying a certain parameter but keeping constant the coupling strength 
$\omega$ between system and environment (if it is sufficiently strong,
$\omega \approx |\gamma_i/2|$). 
The cross coupling between
the two modes vanishes at smaller values $\omega$ where two separated modes 
can be seen in the cross section. The first case is the scenario of
width bifurcation beyond the EP while the second one corresponds to
level repulsion below the EP.

Recently, the transition
from Autler-Townes-splitting (ATS) to electromagnetically induced
transparency (EIT) is considered  theoretically \cite{anisimov} as
well as experimentally \cite{giner}. Two very different processes  
are examined for the case that the
transparency of an initially absorbing medium for a probe field is 
increased by means of a control field.  The explanation is based
on the absorption in a $\Lambda$-type configuration. At very low
control intensity, EIT  occurs while ATS corresponds to the appearance
of two dressed states at large control intensity. The aim of \cite{anisimov}
and \cite{giner} is to discriminate between these two phenomena.

Let us translate our formalism to the language of a $\Lambda$-type system
with the following three states:  the ground state
$|g\rangle$, the excited state $|e\rangle$ and the state  $|s\rangle$
driven by the control field. The detuning $\delta$ from resonance 
provides the energy dependence of the cross section (see
Eq. (\ref{smatr}) for the $S$ matrix) where the two states $|g\rangle$
and $|e\rangle$ appear as resonances at the energies $E_g$ and
$E_e$. The  widths of the two resonance states are $\Gamma_g$ and
$\Gamma_e$, respectively. These values are obtained theoretically from
the complex eigenvalues $\ce_i\equiv E_i+i/2~\Gamma_i$ 
of the non-Hermitian operator $\ch^{(2)}$, see Eq. (\ref{eiv1}).
Not only the width $\Gamma_e$ of  the state  $|e\rangle$ is different
from zero but also the width $\Gamma_g$ of 
the state  $|g\rangle$ is  non-vanishing,  due to the
coupling to the common environment which causes the self-energy to be 
complex (if the coupling is sufficiently strong). This fact
corresponds to the result obtained in \cite{svid} that the 
emitted radiation has a large frequency shift when the size of the
atomic cloud is small compared with the radiation wavelength;
see also the experimental results \cite{roehlsberger,roehlsbergerfdp}. 
We arrive therefore at the following picture. 

The strong control field corresponds to a weak
coupling $\omega$ between system and environment, since the state
$|e\rangle$  may decay  to both the ground state  $|g\rangle$
and the state $|s\rangle$. The interaction $\omega$ is real, predominantly,
and the two states repel each other in energy.

A weak control field, however, is in accordance with a strong coupling
strength $\omega$, since the state  $|e\rangle$ can decay to   
the ground state  $|g\rangle$ without any essential competition with
another decay channel. In this case, the interaction $\omega$ is
imaginary and width bifurcation occurs beyond the EP (without any
further enhancement of $\omega$ as shown in Sect. \ref{num3}).

Thus, we see level repulsion at a strong
control field while width bifurcation (broad bump with a narrow dip)   
dominates at a weak control field. The first case corresponds to ATS
and the second case to EIT. The transition between these two scenarios
occurs smoothly  (Figs. \ref{fig7} to \ref{fig9}). Below the EP
$\omega_{ik}\to 0 ~(i\ne k)$, and $\omega_{ii} \ne 0$ is the complex Lamb shift.

This result agrees with the conclusion drawn in \cite{scully1}
that Fermi's golden rule does not adequately describe Dicke's superradiance.
It is in  qualitative agreement also with the results obtained
and discussed in \cite{anisimov}
and \cite{giner} where the two phases (EIT model and ATS model,
respectively) are considered separately. The EIT model first dominates
in the low Rabi frequency region. With increasing control Rabi
frequency $\Omega$,  the likelihood of the EIT model
decreases, and the ATS model  dominates for larger  $\Omega$. The 
EIT/ATS model transition takes place in a certain critical area 
of the control frequency $\Omega$. 

Instead of the control Rabi frequency $\Omega$, 
the ratio of Im$(\omega_{ij})$ to Re$(\omega_{ij})$ 
controls the cross section in Figs. \ref{fig7} to \ref{fig9} where
$\omega_{ij}$ is the coupling coefficient between system and environment.
According to the definition in Eq. (\ref{om0def}), $y=1$ corresponds
to an imaginary coupling coefficient $\omega_{ij}, ~i\ne j,$  
and $y=0$ to a real one. In the first case, width bifurcation 
dominates and in the second one level repulsion. The transition
between these two  scenarios  is caused by a singularity (EP), as shown 
and discussed in Sect. \ref{num3}. 

The exact definition of $y$ is insignificant as additional
calculations have shown. Important is only that level repulsion
dominates at small $y$ and width bifurcation at large $y$. In the
first case $r_i \approx 1$,  the two resonance states (including the
complex Lamb shift) do
(almost) not overlap and the eigenfunctions $\Phi_i$ are (almost)
orthogonal to one another. In the second case,  $r_i \to 0$, the two states
overlap and interfere strongly, and the  $\Phi_i$ are related to
one another according to Eq. (\ref{eif5}). 

In contrast to the Rabi frequency $\Omega$,
the parameter $y$ cannot be measured directly in an experiment. It is
a theoretical value that follows from the mathematical description of
quantum systems embedded into the continuum of scattering wavefunctions.
One of the relevant measurable values is width bifurcation. 
Starting from two states with approximately equal widths $\gamma_1
\approx \gamma_2$, the difference between the two widths increases up to 
$|\Gamma_1 - \Gamma_2| \to |\gamma_1 + \gamma_2|$ with $\Gamma_1 \to 
\gamma_1 + \gamma_2$ and $\Gamma_2 \to 0$. At the same time, the system 
becomes almost transparent. This transition happens at a {\it finite}
value of the coupling strength $\omega_{ij}$ (causing cooperative 
emission). 

The variation of the coupling strength $\omega$ between the two-level
system and the environment 
can be achieved not only by means of the control field driving the
state $|s\rangle$ as
discussed above. EIT is observed experimentally also  in the 
cooperative emission \cite{roehlsberger} where a
broad bump with a narrow dip appears.
Also in this case, EIT occurs at strong coupling $\omega$.

\section{Conclusions}
\label{concl}

In the present paper, DPTs in open quantum systems and their 
relation to  singular points (EPs) are considered. We found 
that a DPT may appear at different coupling strengths of the
system to its environment. 
\begin{enumerate}
\item[--]
A DPT may occur at high level density 
by successive enhancement of the width of one of the states at the crossings
with nearby states. Due to width bifurcation, the neighboring states  
decouple (more or less) from the environment.
\item[--]
A DPT appears also in a two-level system 
if the coupling strength to the environment is sufficiently
high. Due to width bifurcation, the width of one of the states
may approach zero.
\end{enumerate}
In both cases, the DPT occurs in a finite parameter range where
$r_i<1$ for the phase rigidity of the participating states $i$.  
The basic process is width bifurcation (caused by the existence of EPs
and the related nonlinear terms in the equations) 
by means of which states with very different lifetime are created. 
Fermi's golden rule holds only below the DPT whereas beyond the DPT we
have an anti-golden rule.
In both cases, the neighborhood of the EPs is non-symmetric
in relation to energy and (or) time (Sect. \ref{disc1}). 

The DPT may explain many results observed experimentally
which are counterintuitive at first glance, e.g. 
\cite{phaselapses,phaselapses2} in a many-level system and 
\cite{past1,past2} in a
two-level system. According to the results of the present paper, 
some features of  Dicke's superradiance may be related also to a DPT.

Due to the relation of a DPT to irreversible processes, it
is interesting  to prove experimentally
the influence of a nearby state onto the symmetry properties of
an EP. Our results suggest that the symmetry is influenced  
by the existence of another state in the 
neighborhood of the EP where  $r_i < 1$ holds for the phase rigidity. Under this
condition, the phases of the eigenfunctions $\Phi_i$ relative to one
another are not fixed and irreversible processes may appear. 

While DPTs are known to appear in open quantum systems
and related (qualitatively) to singular points (EPs) since several years
(references can be found in Sect. 6 of \cite{fortsch1} and in Sect. 5 
of \cite{jmp}, see also \cite{past1,past2,phaselapses,phaselapses2,cnr11}), 
some properties of Dicke's superradiance are  referred to singular points
first in the present paper. The results 
should be proven by further experimental studies. Furthermore,
the results  are surely important  for fundamental questions
of quantum mechanics. Above all however, they are of high value for 
applications where long-lived states
play an important role, e.g. for the storage of information, for quantum
memory  and for quantum filter. 
As shown in the present paper,  Fermi's golden rule is replaced by an
anti-golden rule under certain conditions. Here,
states with a long lifetime may occur due to width
bifurcation in a two-level system at a coupling strength $\omega$ 
that is large and causes coherent emission, but is too small
in order to destroy the whole system. In such a case, 
the wavefunction of the long-lived (subradiant) state is well defined.

\begin{appendix}
\section{Dynamical phase transitions}
\label{ap1}

The notation {\it dynamical phase transition (DPT)} 
is used in literature in order to describe the 
redistribution of the spectroscopic
properties of an open quantum system under critical conditions
when the system's properties are controlled by a parameter, see
e.g. the review \cite{top}. 
The redistribution is mostly counterintuitive and is related to a
violation of Fermi's golden rule. 
Two examples are the following. (i) A DPT is observed experimentally 
and explained theoretically in the spin swapping operation \cite{past1,past2}. 
While Fermi's golden rule holds below the DPT, it is violated above
it. (ii) The experimentally observed so-called phase lapses 
in quantum dots \cite{phaselapses} can be explained by
means of a DPT occurring due to the non-Hermiticity of the Hamilton operator 
\cite{phaselapses2} and is related to
width bifurcation caused by exceptional points (EPs)  \cite{elro2}.

The Hamiltonian of an open quantum
system is non-Hermitian due to its coupling to the environment of
scattering wavefunctions into which the system is embedded. The
eigenvalues $\ce_i \equiv E_i + \frac{i}{2} \Gamma_i$ (with $\Gamma_i \le 0$)  
of the non-Hermitian Hamiltonian are complex and provide
not only the energies of the states but also their lifetimes (inverse
proportional to the widths $|\Gamma_i|$). At several points, 
two eigenvalues may coalesce. These points  
are singular and called mostly {\it exceptional points (EP)} (after Kato
\cite{kato}). Mathematically,  DPTs are related to EPs  around which  
the phase rigidity (\ref{eif11}) is reduced. 
It is possible therefore that one of the states of the open system
aligns to the 
scattering states of the environment and becomes short-lived  
while the other states decouple from the environment and become long-lived.   
Physically, DPTs are the result of selforganization 
\cite{selforg} and occur under the influence of the
environment in which the system is embedded. Most important is
the formation of different time scales in the open quantum system 
at parameter values beyond the critical ones, see e.g. \cite{fortsch1}.

A non-trivial question is whether the DPTs with properties sketched
above have something in common with a phase transition in the sense of
thermodynamics. This question is studied  some years ago \cite{jumuro}. 
Under critical conditions, a reorganization of the spectrum takes
place that creates a bifurcation of the time scales with respect to
the lifetimes of the resonance states. The conditions under which the
reorganization process can be understood as a second-order phase
transition are derived analytically and the results are illustrated by
numerical investigations. The conditions are fulfilled, e.g., for a
uniform picket-fence distribution with equal coupling of the states to
the continuum. Other examples are also considered. In all cases, the
reorganization of the spectrum occurs at the critical value of the
control parameter {\it globally} over the whole energy range of the
spectrum. All states act {\it cooperatively}. The length scale
diverges as well as the degree of non-Hermiticity of the Hamiltonian
(which is related to the phase rigidity defined in (\ref{eif11})).
Both values diverge at the EP even in the two-level
case if the coupling to the continuum is sufficiently strong
(see also Figs. \ref{fig3}.c, \ref{fig7}.e and \ref{fig8}.e).
The time scale of an open quantum system \cite{fortsch1}
differs however from that of a closed system the Hamiltonian of which is
Hermitian. Further, observables such as the cross section  behave smoothly
at the critical value of the control parameter due to the
bi-orthogonality of the eigenfunctions of a non-Hermitian operator
involved in the S-matrix (see, e.g., \cite{top} and the numerical
results shown in Sect. \ref{num3}). 

Phase transitions in open quantum systems which are
associated with the formation of long-lived and short-lived states
according to  \cite{jumuro}, are related to EPs first in
\cite{hemuro}. More recent results can be found in, e.g., \cite{top}
and \cite{fortsch1}.  Most important
difference between Hermitian and non-Hermitian quantum mechanics
is surely the fact that the phases of the eigenstates (relative to one another) 
of a non-Hermitian operator are not rigid in the neighborhood of an
EP (see (\ref{eif11}) and the corresponding discussion), 
while they are rigid everywhere in the standard Hermitian quantum physics. 

It should be mentioned here, that a DPT cannot be described by using a
master equation. The reason is the basic assumption for the description of
the dynamics of a quantum system by a master equation,
according to which  the time scale $\delta t$  has to satisfy the condition 
$\delta t \ll\tau_S$ where $\tau_S$ is the characteristic
timescale of the system. 
This condition is not fulfilled near to an EP since this point  is 
a crossing point of two eigenvalue trajectories. Near to an EP,
the characteristic time scale $\tau_S$ approaches therefore zero 
(see also (\ref{eif5}) 
for the two wavefunctions at the EP). This fact does surely not play any
role for an isolated EP in a many-level system. It does
however not allow to describe the many-level system by a master equation in the
regime of many dense lying EPs, i.e. when a DPT is approached.

Another question is the relation of  DPTs as defined above, to
dissipative phase transitions studied in, e.g., \cite{diss1}. Dissipative
phase transitions occur in open quantum systems when the 
asymptotic decay rate (defined by the 
smallest linewidth after the width bifurcation)  vanishes. 
According to the above discussion, this case is a special DPT occurring
when (at least) one of the long-lived states decouples completely 
from the environment and a non-analytical steady state appears at the
critical (exceptional) point. An example is considered numerically 
in Sect. \ref{num3} and is discussed in Sect. \ref{disc2}.
The numerical results obtained for dissipative phase transitions
\cite{diss2,domo1,domo3,domo2,diss1}
show the same characteristic features as those received in the present model,
Figs. \ref{fig1}.a,b and, above all, Figs. \ref{fig3}.a,b. 
The numerical results obtained for DPTs prove the analytical results 
(\ref{int6b}) to 
(\ref{int6d}) and are very robust. Furthermore, the results of both
approaches are related to selforganization by the authors, 
respectively \cite{domo3,domo2} and \cite{selforg}.

It should be added here that DPTs in the sense of the definition
given above are found and discussed in different real systems
theoretically as well as experimentally.  
Two examples are mentioned above, \cite{past1,past2} and
\cite{phaselapses,phaselapses2}. In the first case, the DPT causes a
violation of  Fermi's golden rule. In the second case,
experimental results which could not be explained in the framework of
standard Hermitian theory in spite of much effort, could be interpreted by means
of a DPT  occurring due to the non-Hermiticity of the Hamilton operator
in the open quantum system. 
Further examples are, among others, the correlated behavior of
conductance and phase rigidity in the transition from the weak-coupling to the
strong-coupling regime in quantum dots \cite{burosa} and 
laser-induced resonance trapping in atoms \cite{marost,marost03}.
Some years ago, a direct experimental proof of a DPT is performed in an open 
microwave cavity \cite{stm}. In the present paper, generic results on
DPTs occurring in open quantum systems are studied and compared with the 
experience obtained from the study of real physical systems. To this
end, a toy model is used.

\section{Open quantum system}
\label{ap2}

An open quantum system is embedded into an environment. 
The natural environment is the
continuum of scattering wavefunctions that always exists. It can 
be modified, however not deleted. The Hamilton operator ${\cal H}$
of the system is non-Hermitian. Its eigenvalues contain the
information on the interaction of states via the environment, 
i.e. the feedback between system and environment. 

The Feshbach projection operator technique \cite{feshbach} containing
the non-Hermitian Hamilton operator  ${\cal H}$,    provides a
convenient method to describe an open many-body quantum system.   
Using this method, first the {\it energy-independent} many-body problem
of the system (with the Hermitian Hamiltonian
$H_B$) is solved in the standard manner.  These solutions provide the 
energies $E_i^B$ and wavefunctions $\Phi_i^B$ of the discrete 
states (eigenvalues and eigenfunctions of $H_B$)
with inclusion of the so-called  {\it internal} interaction.
In a second step, the {\it energy-dependent} scattering wavefunctions
$\xi_c^E$ of the  
environment are calculated and, further, the (energy-dependent) 
coupling matrix elements  
\begin{eqnarray}
\gamma_{k c}^0  = 
\sqrt{2\pi}\, \langle \Phi_k^B| V | \xi^{E}_{c}
\rangle 
\label{form3}
\end{eqnarray}
between the discrete states of the system and the environment 
are evaluated (see \cite{top}, Sect. 2.1). 
The non-Hermitian Hamiltonian ${\cal H}$ contains the selfenergy of
the states in the diagonal matrix elements and  
the interaction of the different states via
the environment (the so-called {\it external} interaction) 
in the nondiagonal matrix elements \cite{ro01}. The Hamiltonian $\ch$ reads
\begin{eqnarray}
\ch\; =\; H_{B} + V_{BC} G_C^{(+)} V_{CB} 
\label{form5}
\end{eqnarray}
where $V_{BC}$ and $V_{CB}$ stand for the coupling between system and
environment and
$G_C^{(+)}$ is the Green function in the subspace of scattering states.
The non-Hermitian operator $\ch$ can be diagonalized: the eigenvalues
$\ce_i$ are complex and the eigenfunctions $\Phi_i$ are complex and
biorthogonal. 

The external interaction of the states via the continuum 
is complex, generally. The principal value integral is 
\begin{eqnarray}
{\rm Re}\; 
\langle \Phi_i^{B} | \ch |  \Phi_j^{B} \rangle 
 -  E_i^B \delta_{ij} 
=\frac{1}{2\pi} \sum_{c=1}^C {\cal P} 
\int_{\epsilon_c}^{\epsilon_{c}'} 
 {\rm d} E' \;  
\frac{\gamma_{ic}^0 \gamma_{jc}^0}{E-E'} 
\label{form11}
\end{eqnarray}
and the residuum reads
\begin{eqnarray}
{\rm Im}\; \langle \Phi_i^{B} | \ch |
  \Phi_j^{B} \rangle =
- \frac{1}{2}\; \sum_{c=1}^C  \gamma_{ic}^0 \gamma_{jc}^0 
\label{form12}
\end{eqnarray}
where $C$ is the number of continua. It is $C=1$ when the different
states are coupled to one common continuum (notation $\gamma_{ic}^0 
\to \gamma_{i}^0$ usually by ignoring the index $c$). 
The interaction of the states of the system with and  via the environment 
is involved in the eigenvalues ${\cal E}_i$ and eigenfunctions 
$\Phi_i$ of the Hamiltonian  ${\cal H}$.
It is therefore relatively easy, in this formalism, 
to study the influence of the environment onto the states of the system.

According to (\ref{form5}), the
matrix elements of $\ch$ consist {\it formally}
of a first-order and a second-order interaction term.
The first-order term stands for the direct ({\it internal}) interaction 
$V_{ij}$ between the two states $i$ and $j$ which is involved in $H_B$. The
states with inclusion of the internal interaction are called usually  
dressed states. The second-order term describes the ({\it external})
interaction $\omega_{ij} \propto V_{ic}\, G_c\, V_{cj}$ via the continuum
$c$ (where  $G_c$ is the Green function in the  
continuum of scattering states).  The second-order term 
consists of the principal value integral (\ref{form11}) 
and the residuum(\ref{form12}). It is
therefore complex, generally. Although it is of second order, it
determines, under certain conditions, the dynamics of the system. 
Furthermore, $\omega_{ii}$ contains the interaction of 
the state $i$ with the environment, i.e.  the {\it self-energy}
of the state which is analog to the Lamb shift known in atomic
physics.  

In order to study the interaction of two states via the common
environment it is convenient to start from resonance states coupled to
the continuum, instead from discrete states. We  use the matrix
representation of $\ch$ and  consider, as examples, the symmetric
$2 \times 2$ matrix (\ref{int3})
for 2 states and the $N \times N$ matrix (\ref{form1}) 
for $N>2$ states. The physical meaning of the matrix elements follows
from equations (\ref{form3}) to (\ref{form12}).

\end{appendix}


\begin{thebibliography}{99}
\bibitem{comment2} In difference to \cite{fortsch2,elro2}, the definitions
  $\epsilon_i=e_i+\frac{i}{2}\gamma_i$ and ${\cal E}_i=E_i+\frac{i}{2}\Gamma_i$
with $\gamma_i\le 0$ and $\Gamma_i\le 0$ are used in the present paper.
\bibitem{comment} In studies of other researchers, the factor $i$ in 
(\ref{eif5}) does not appear. This difference is discussed 
and compared with experimental data in the Appendix of \cite{fortsch1} and
in Sect. 2.5 of \cite{top}.
\bibitem{simulation}  The simulation rests on the doorway picture. 
The $N$ states with the widths $\gamma_{i}^0$ can be simulated by 
one doorway state $N=d$ with  large width $\gamma_{d}$ and $N-1$ states
with  small widths $\gamma_{i}^{N-1}$. The width of the doorway state is
large due to its  strong coupling to the common continuum.
In any case, the number of parameters is $2N$ ($N$ for the widths and
$N$ for the energies of the $N$ states, see \cite{mahaux}).
\end{thebibliography}
\end{document}